\documentstyle[prl,multicol,aps,epsf]{revtex}
\begin{document}
\title{Quantum Spin Nematic States in  Bose-Einstein Condensates }
\author{Fei Zhou}
\address{ITP, Utrecht University,
Minnaert building, Leuvenlaan 4, 3584 CE Utrecht, The 
Netherlands}
\date{\today}
\maketitle

\begin{abstract} 
We review some recent results on discrete symmetries and
topological order in spinor Bose-Einstein condensates (BECs) of 
$^{23}Na$.
For spin one bosons with two-body scatterings dominated by a total spin
equal to two channel, the BECs are in quantum spin nematic states at a low
density limit. 
We study spin correlations in condensates at different limits and analyze hidden 
symmetries using a non-perturbative approach developed recently.
We further more investigate the influence of
hidden $Z_2$ symmetries and $U(1)$ quantum orders in
symmetry partially retored states, particularly the effects on topological 
excitations.
\end{abstract}

\begin{multicols}{2}

\section{introduction}

BECs of alkali atoms first created and studied experimentally in 1995-1997
revived theorists' interest on this 
subject\cite{Anderson95,Davis95,Bradley97}.
Because the Zeemann coupling in general provides 
stronger
confining potentials for atoms compared with other trapping methods, and  
a rf evaporative cooling is needed to achieve BECs,
the first generation of BECs was realized in magnetic traps in
laboratories.
In those early experiments, spins of atoms were completely polarized and 
BECs with single component were achieved, similar to the BEC of $^{4}He$ 
studied a few decades ago. Apart from the complication of
the confining potentials, the
theoretical description of BECs of alkali atoms with positive scattering 
lengths in magnetic traps can be carried out 
in the Gross-Pitaveskii approach\cite{Nozieres,Griffin95}.
The activities on this subject are enormous and I should refer
to an excellent review by Dalfovo et al.\cite{Pitaevskii}.

The physical properties of single component BECs of alkali atoms
depend on two length scales. The interatomic distance,
which is about a few hundred nano-meters in most of experiments,
characterizes the kinetic energy in the problem. The BEC transition 
temperature which is determined by this length scale is about
a few micro Kelvins or lower in experiments. The upper limit of
the atomic density in BECs and the transition
temperature are imposed by three-body 
recombination
processes which lead to trap losses in experiments\cite{Ketterle}.
The second length scale is the two-body scattering length, which is about
a few nanometers for alkali atoms.
It characterizes the interactions between atoms and therefore the chemical
potentials. Given the density of atoms and the scattering lengths in most 
of experiments, the chemical potentials are typically  
of the order of a few hundreds nano Kelvins.
Amazingly, the scattering length depends on magnetic fields and
the variation in scattering
lengths is particularly substantial when the atoms are 
close to the Feshbach resonance\cite{Stwalley,Tiesinga}. 
This remarkable property of alkali atoms was observed in $Na$
by Inouye et al.\cite{Inouye98}, in $Rb$ by Courteille et al.
\cite{Courteille},
Roberts et al. \cite{Roberts} and recently Cornish et al.\cite{Cornish}, 
in $Cs$ by Vuletic et 
al.\cite{Vuletic}.
Collapsing  of BECs of $^{7}Li$ atoms with negative scattering lengths 
was studied by Sackett et al.\cite{Sackett}. Dynamics of
collapsing was investigated theoretically in
\cite{Duine00} and remains to be fully understood.

Correspondingly,
there are at least two classes phenomena investigated very extensively 
by experimentalists. The first one is connected with  
quantum coherence of BECs. Quantum interference between two condensates
(both spatial and temporal) and the creation of vortices belong to this 
category\cite{Andrews97a,Anderson98,Matthews99,Madison00,Chevy00}.
For instance, vortices were first created using 
a combination of internal transition and external
rotation by Matthews\cite{Matthews99}, as theoretically 
suggested by Williams and Holland in 1999. 
Arrays of vortices were found by Madison et al.\cite{Madison00}, who
stirred BECs by rotating the trap.
Superfluidity of BECs has been examined by
Raman et al.\cite{Raman00} who demonstrated frictionless flow
and by Marago et al.\cite{Marago00}, who observed beautiful undamped 
scissors mode suggested by Guery-Odelin and Stringari in 
1999\cite{Guery-Odelin99}.

The second class is related with scatterings between atoms, or 
the collective
behavior of BECs.
The AC Josephson effect of two internal states with different scattering
lengths and zero sound excitations at coherent limit fall into this 
category\cite{Hall98,Stamper-Kurn98,Mewes,Andrews97b}.
The second class phenomena are of particular interest from condensed 
matter physics points of view
and let me describe two interesting experiments in some details.
In the experiment in \cite{Hall98}, Hall et al. first prepared
$^{87}Rb$ in a hyperfine spin state $|F=1, m_F=-1>$ in magnetic traps.
By applying external microwaves or so called $\frac{\pi}{2}$-pulses, they 
successfully transformed about fifty percent of atoms from $|1,-1>$ state 
into $|2,-2>$ state. The chemical potentials of these two condensates
are different because of
different scattering lengths. 
After certain waiting time $T$,
a fraction of the condensate at $|2,-2>$ state is transformed back
to $|1,-1>$ state by another $\frac{\pi}{2}$-pulse. The interference 
between the 
original condensate and the newly converted condensate depends on the 
accumulated phases of two condensates after time $T$, which
are different because of two different chemical potentials
$\mu_i=4\pi a_i\rho/M$ for condensates with a density $\rho$,
a scattering length $a_i$ and an atomic mass $M$.
The temporal dependence of the interference observed by Hall et al.
does exhibit periodical oscillations at $ms$ scale, expected by
a simple estimate of the chemical potentials.

The observation of zero sound excitations in $^{23} Na$ BECs was carried
out by Stamper-Kurn et al.\cite{Stamper-Kurn98} in 1998. After a temporal 
magnetic 
field was applied to disturb the BEC, a breathing of condensate
was observed at a temperature much lower than the BEC transition 
temperature. Snap shots taken at $5ms$ apart indicate
oscillations in the size of the BEC with a period of a few $ms$, 
corresponding 
to the time interval for a sound like excitation to travel from
one end of the BEC to the other one.
The sound velocity, being proportional to the square root of
the compressibility, depends on the scattering lengths and appears
to be consistent with the frequency of the oscillations.
These oscillations were also shown quickly damped at the temperature
close to the transition point, caused by scatterings with a thermal cloud.

BECs with multiple components were created and 
studied a few years later\cite{Myatt,Stamper,Stenger}.
Particularly, Stamper-Kurn et al. successfully used optical beams
to trap $^{23}Na$, which confines alkali 
atoms independent of 
spins and liberates spin degrees of freedom\cite{Stamper}. 
For alkali atoms $^{23}Na$ or $^{87}Rb$, with nuclear spin
$I=3/2$ and electrons at $s$ orbits, 
the energy splitting between hyperfine multiplets is of 
order
$100mk$. At temperatures as low as a few hundred nano Kelvin, $^{23}Na$
and $^{87}Rb$ atoms can be considered as simple bosons with a hyperfine
spin $F=1$. 
Liberation of spins in BECs in an optical environment indeed 
leads to a new dimension of studying BECs, especially
spin correlated condensates.
This further inspired many theoretical 
activities on spinor
Bose-Einstein condensates\cite{Ho,Ohmi,Law}. 
The spin correlated states in $^{23}Na$ were afterwards studied 
experimentally by 
Stenger et al.\cite{Stenger}.

The new complication in BECs in optical traps surely 
arises from the extra internal degree of hyperfine spins.
When $N_{m_F}$( $m_F=0, \pm 1$) atoms condense at each of the 
hyperfine
states $|F, m_F>$ but all condense at the lowest orbital 
state, the ground state of $N(=\sum_{m_F} N_{m_F})$ 
noninteracting spin $F=1$ 
bosons has a ${\cal D}_s$--fold spin degeneracy,

\begin{equation}
{\cal D}_s=\frac{(N+2)(N+1)}{2},
\end{equation}
by contrast to its magnetically trapped 
cousins. Presumably, hyperfine spin-dependent scatterings lift
the degeneracy and lead to spin correlated states. 
Optically trapped BECs therefore set up a platform
for the study of the quantum magnetism in many boson systems and
add a new twist to already extremely fascinating 
systems.

For ultracold atoms in BECs,
the spin-dependent two
-body interaction is characterized by
$U_2({\bf r_1}, {\bf r_2})=\delta({\bf r}_1-{\bf r}_2)[c_0 + 
c_2 {\bf F_1} \dot {\bf F_2}]$, as suggested in an early paper\cite{Ho}. 
Here $c_0=(g_0+2g_2)/3$, $c_2=(g_2 -g_0)/3$;
$g_F=4\pi \hbar^2 a_F/M$, M is the mass of the atom and
$a_F$ is the s-wave scattering length in the total spin $F$ channel.  
Thus,
spin correlation in BECs is determined by $c_2$. For
$^{87}Rb$ studied experimentally by the JILA group\cite{Myatt}, $g_2 < 
g_0$ or $c_2 <0$
and the scattering
is dominated by the total spin $F=0$ channel.
In the ground state, all spins of atoms prefer to 
align in a certain direction and have a maxima magnetization\cite{Ho}.
For $^{23}Na$ studied experimentally by the MIT group\cite{Stamper}, 
the scattering between $^{23}Na$ atoms is dominated by 
the total spin $F=2$ channel, i.e.
$g_2 > g_0$.
The scatterings between $^{23}Na$ atoms thus lead to 
"antiferromagnetic" spin correlation.
In this article, I will fully explore the properties of the BEC of 
$^{23}Na$.

Efforts have been made to understand
the ground state properties, exact excitation spectra, collective modes
and topological excitations.
Theoretically, issues
of spin ordering in the spinor BECs are of particular interest
and were treated by Ho, and also Ohmi and Machinda in 1998.
In both works, the spin ordering was discussed in a mean field
approximation, in the context of the Gross-Pitaveskii equation. Ho also 
obtained 
two branches spin wave excitations by a Gaussian expansion.

The first attempt to understand spin correlation beyond the mean field
limit was made by Law, Pu and Bigelow\cite{Law}. Law et al. observed that
the microscopic Hamiltonian of N spin-1 Bosons interacting via
a spin-dependent two-body potential given above commutes with the total
spin of the system; the
spin is a good quantum number and the exact eigenstates should be
eigenstates
of the total spin operator. 
When $c_2 > 0$, the ground state should be a spin singlet
and rotationally invariant.  
This was explicitly demonstrated in a zero mode
approximation, where the finite momentum sector of the Hilbert
space was neglected\cite{Law}. 
Using a four-wave mixing method, Law et al. obtained
the exact excitation spectrum in a $0D$ limit, a limit
where all spatial fluctuations are neglected.
These results in $0D$ case were also derived by Castin and 
Herzog\cite{Castin}, using a geometrical description very close to the 
one I am going to introduce in this paper. 
Law et al.' s results, though are barely relevant to the particular
experiment done by Stenger et al., given the number of atoms,
time scales during which the experiment was performed and 
small biased magnetic fields in the experiment, nevertheless
inexplicitly reveal the role of symmetry restoring processes and
the nature of quantum fluctuations of the spin order in the problem.
The even-odd effect in the excitation
spectrum observed by Law et al. further 
implies a discrete symmetry
in the problem which I believe has a profound 
origin.

Ho and Yip later found that the singlet states obtained by
Law et al. are rather 
sensitive to an external magnetic field\cite{Ho-Yip}. In a thermal 
dynamical limit, 
an infinitesimal magnetic field will stablize
the symmetry broken state and the rotationally invariant state
is irrelevant as far as experiments with a small magnetic field
are concerned.
This feature of $^{23}Na$ was later realized to relate with
the level spacing of the low lying excitations\cite{Zhou1,Zhou2}. For 
the $0D$ 
dimensional BEC, the energy of low lying excitations scales down as the 
number
of atoms and the quasi-degeneracy makes the system ultrasensitive to
magnetic fields. 
A symmetry broken state, which can be constructed out of low lying
excitations has an energy inversely proportional to the number of
atoms involved and can be pinned by an infinitesimal field when the number 
of atoms is infinity.

Topological excitations in the spinor BECs of $^{23}Na$ have been paid a
special tribute to by a few groups. Linear defects were studied by Leonhart
and Volovik\cite{Volovik00}, who pointed out the existence of Alice
strings in the BECs of $^{23}Na$. Textures of the Skyrmion type were
pointed out in\cite{Ho}; and most recently, point defects were
examined\cite{Stoof}. Except the work of Leonhart and Volovik, most of the
works on this subject were based on Ho's original identification of the
internal space, which missed the important discrete symmetry of $Z_2$
type. Therefore, many new aspects of topological defects studied in this
paper have never been investigated.  For instance, in a recent
work \cite{Stoof}, the authors applied the nonlinear $\sigma$ model
derived in \cite{Zhou1,Zhou2} to study static and dynamical
properties of hedgehogs. Unfortunately, the authors of that work failed to
observe the correct internal space. Many aspects of the point defects in
the BECs of $^{23}Na$, particularly those under the influence of the
entanglement of the phase and spin orders including 
the homotopical indistinguishability between $\pm$ hedgehogs,
between hedgehogs and closed loop $\pi$-disclinations were previously
discussed \cite{Zhou1}. I will review those results in
this article.

The pursue of having a nonperturbative approach beyond a
zero dimension approximation was first made in\cite{Zhou1,Zhou2}.
That approach allows one to identify that the low energy spin dynamics in 
the BECs of spin one atoms with antiferromagnetic interactions belongs to
the same universality class of the nonlinear-$\sigma$ model(NL$\sigma$M), 
a relatively
well known subject, thanks to many studies in the context of field theory.
The NL$\sigma$M also became known to many condensed matter physicists 
working with antiferromagnetic systems after a paper in 
1983\cite{Haldane83}.

In \cite{Zhou1,Zhou2}, 
the spin dynamics in the problem of N spin one bosons
interacting via a spin-dependent two-body potential with $c_2 >0$
was mapped into an $o(n)$ nonlinear sigma model (NL$\sigma M$). $n=3$
at the zero magnetic field limit and
$n=2$ in the presence of a weak magnetic field.
With the help of this mapping, the many-body aspect
of $^{23}Na$ becomes rather explicit and many properties including
spin correlations in $1d$ can be studied.
Furthermore, energetics and dynamics of topological excitations 
in BECs can therefore be easily analyzed in this approach.

At a low density limit, it is also found that the ground 
state 
has a similar discrete symmetry as that of
a uniaxial nematic phase of liquid
crystals, except a $\pi$-phase factor as discussed extensively in 
\cite{Zhou1,Zhou2} 
and for this reason the condensate is at a quantum spin nematic
state({\em QSNS}). 
The ground state degeneracy space is $[U(1)\times
S^n]/Z_2$, with $n$
depending on a magnetic field.
The antiferromagnetic scatterings between cold atoms lead to a $Z_2$ 
symmetry breaking.

This discrete symmetry was later appreciated by Demler et al.
in the context of low lying excitations\cite{Zhou3,Zhou4}.
Demler et al. noticed that the $Z_2$ symmetry in a symmetry broken state 
actually
also appears in all low lying states.
By imposing a proper projection in a functional integral approach,
they showed that the full low energy physics in the phase and spin sector
is given by $Z_2$ gauge fields coupled with two matter fields 
characterized by
an {\em XY} model and a NL$\sigma$M.
This work as well as the previous work \cite{Zhou1,Zhou2} open a new door 
for the studies of many-boson states.
Many issues, such as fractionalization of topological excitations,
fractionalization of quasiparticles and topological order
were studied and investigated with the help of this 
mapping\cite{Zhou5,Zhou6}.
A number of new predictions have been made for the spinor BEC, most of 
which have
never been studied experimentally.
The mapping also provides a starting point for the future study of some
strongly correlated fermionic systems.

The purpose of this article
is to provide a guideline and introduction to some ideas on the spinor 
BECs
developed very recently in \cite{Zhou1,Zhou2}.
Through out the article, I will emphasis on the concepts of topology,
at levels of both the mean field approximation, which is suitable for a 
discussion
on symmetry broken states and beyond, which is essential for
the study of symmetry restored states.
This point of view 
might be dramatically different from the conventional ones shared by 
atomic
physicists. On the other hand, it seems to me that the nature has
chosen to behave in such a unconventional, complicated way
that I am left with no alternatives, or a simpler 
presentation. It also appears to my collaborators and me
during this investigation that this is a subject full of surprises
and new issues keep on poping up. At the time of writing,
I present results which I believe to be relevant to 
BECs investigated in experiments. There are many other issues still 
remaining to be explored.
Furthermore, some descriptions given here are qualitative and conceptional,
so as to be communicated in a most efficient way.
Like many other new subjects, a more quantitative aspect of some 
properties can be achieved only after basic notions are established and 
will be present 
in a series of separate works.  
At last, I should point out that many strongly correlated
electronic systems are believed to share very similar topological 
properties\cite{Senthil,Demler01} and therefore the BECs might be systems 
where some
of basic ideas established in strongly correlated electron systems can be 
tested experimentally.
Understanding of the spinor BEC might as well
shed some light on other Fermionic superfluids such as
p-wave triplet superconductors.

So, I will review the nonperturbative approach recently formulated.
Results obtained 
previously by different groups
with other methods can be reproduced in this new representation.
I will of course go beyond previous framework and outline
many novel results obtained in the last two years
\cite{Zhou1,Zhou2,Zhou3,Zhou4,Zhou5,Zhou6}. 
In section II,  a mapping from the original microscopic model of N
spin one Bosons
with $c_2 >0$ into  a $NL\sigma M$ at long wave length limit
will be introduced. Furthermore,
the corresponding N-body wave function is shown to be invariant 
under a
$180^0$ rotation of a vector ${\bf n}$ and a $U(1)$ gauge transformation.
At last, $Z_2$ gauge fields are introduced in the model to 
enforce the discrete symmetry.

In section III, symmetry broken states are examined.
The internal order parameter space for
the BEC is identified as $[S^1\times S^2]/Z_2$(zero field limit).
Linear, point defects and "particle like" textures
are discussed with an emphasis on the influence of the $S^2$ group and
the discrete $Z_2$ symmetry.
Particularly,  a family of strings representing
$\pi$ spin
disclinations superimposed by superfluid vortices of half integer 
circulation are studied.
The distinguishabilities of $\pm$ hedgehogs, both physical and
homotopical, are analysised.
The connection between closed loop $\pi$-disclinations and hedgehogs
is established.

The effect of an external
magnetic field is also reviewed.
The NL$\sigma$M in this case has
two components instead of three at zero external field.
The ground state is degenerate in a manifold 
$[U(1)\times S^1]/Z_2$.

In section IV, 
symmetry unbroken states 
are studied 
in the scheme of the NL$\sigma$M.
The possibility of having
nematic order-disorder phase transitions and the nature of
spin correlation are addressed.
The ground state of N spin one Bosons
interacting via $c_2>0$ spin-dependent two-body
potentials in 1d limit is also present. 
In section V, the influence of the $Z_2$ symmetry on the symmetry unbroken 
states is discussed.
One of the most interesting consequencies of having $Z_2$
symmetries is the possible fractionalization of
topological excitations.
Under the influence of the $Z_2$ symmetry, it is shown
that vortices can be fractionalized into
half-vortices. It is also pointed out that instead of 
a condensate of individual atoms, one can have a singlet pair 
condensate, which supports free $Z_2$ vortices.

In section VI, the property of connection fields
is analysised and hidden topological order is identified 
in spin disordered BECs.
The existence "topological order" is argued due to 
an "order from disorder" mechanism.

In section VII, I discuss the symmetry restoring of a finite size
nematically ordered state. Quantum mechanical nature of the order 
parameters in finite size $^{23}Na$ BEC was addressed
in connections of experimental probes,
taking into account the metastability of the atomic gas. In
section VIII, I make some remarks on the relation between 
the BECs considered here and other strongly correlated magnetic systems.

\section{Geometric description}

The inconsistency between the mean field point of view
taken in \cite{Ho,Ohmi} and the spin singlet ground state obtained 
in \cite{Law},
lies in the heart of the quantum symmetry restoring.
If one starts with the symmetry broken
solution in the GPE approach,
for instance, one can ask how the rotation symmetry is restored after all
and when that happens.
A more practical question would be,
given the life time of the metastable
BEC gaseous cloud studied in the experiment, is the rotationally invariant
state relevant?
This question motivates us to develop a nonperturbative approach.
The answer to this question requires a complete characterization of
nonlinear spin dynamics.

The second issue  
concerning us is the validity of the zero mode approximation\cite{Law}.
Though the exact solutions in a zero mode approximation
provide information of
low lying excitations, they 
fail to yield spatial-temporal spin fluctuations,
which are essential for the most general description 
of the ground state of $N$ interacting $F=1$ Bosons.
The finite momentum sector of Hilbert space
is clearly critical for such a discussion.

It is also worth commenting that
the GPE approach with three components\cite{Ho,Ohmi},
on the other hand, 
does permit a gradient expansion around the symmetry
broken solution
in principal. However, such a procedure is feasible
only when the derivation of the true ground
state from the uniform solution is small. 
Practically, the GPE approach works
in a small vicinity of the symmetry broken solution
in the order parameter space and is far from being a general
description.

The third and perhaps the most important one is the
description of nonlinear spin dynamics in the spinor BEC.
The failure of the generalized 
GPE approach far away from its mean field solution is
largely due to the fact that
the collective
variables chosen there are not suitable for a
description of the nonlinear spin dynamics.
For this reason, in the framework of the GPE approach,
the interactions between low lying excitations(spin waves)
which eventually disrupt the symmetry broken state are barely 
tractable.

For instance, when the zero point rotation energy of each atom
caused by two-body scatterings is high enough, long-range spin
order predicted in the GPE approach is destroyed even at zero
temperature(see details in section IV).
In that case, 
spin correlations 
can be nontrivial and yet not characterized by the simple
solutions of the GPE.
In addition, the spin correlated state in a very anisotropic
trap, or $1d$, is clearly beyond the validity of the GPE
approach.


Efforts along this line had been carried out recently.
To have a  general approach of studying 
spin correlated states of spin one bosons with $c_2 >0$
was the purpose of those works.
It is
reviewed in the rest of this section.
I will demonstrate a geometrical approach which allows
us to go far beyond the three-component Gross-Pitaveskii 
approach and explore a variety of possible spin correlated states.

\subsection{NL$\sigma$M characterization of spin dynamics}

Following the discussions above,
to study the spin correlated BEC,
it is important that 
a set of collective coordinates are correctly
defined. Idealistically,
these variables should be chosen to 
a) completely characterize the nonlinear spin dynamics in
the whole parameter space and b) to form a simplest
representation for two-body spin dependent scatterings. 
For this purpose, we use two vectors
$({\bf \Omega}_1,{\bf \Omega}_2)$ living on two 
unit spheres to describe the nonlinear spin dynamics 
in the presence of spin-dependent scatterings.
We are going to map the N interacting spin one Boson
problem into an $O(3)$ NL$\sigma$M self-consistently.

We make two assumptions to proceed further.
First, there exists a class of spatially varying
N-body wave functions,
out of which the N interacting spin-1 Bosons 
ground state and the low lying 
excitations can be constructed.
The whole class is characterized by a few collective 
coordinates( ${\bf \Omega}_{1,2}$ introduced below).

Second, the energy of this class N-body wave functions under consideration
can be expressed 
in terms of a finite number of collective variables.
Quantization of the collective modes
yields elementary low lying excitations,
which can interact with each other strongly. The exact 
N-body wave function of the ground state
is characterized by the zero point
motions of collective variables and their multi-couplings. 
Practically, in most limits, the ground state is conveniently 
characterized by the correlation functions.
The low energy spin excitations are of a collective nature;
the excitations of individual spins are neglected in 
the leading order approximation $O(1/N)$.

The correctness of this approach is justified by the following measure: 
The resultant theory which is written in terms 
of collective variables instead of N-body wave functions of individual
atoms should produce the correct nonlinear spin dynamics,
such as Euler's equation of motion, conservation laws, 
at the long wave length limit.
The theory is self-consistent if this measure is satisfied.
The derivation  of the mapping was obtained   
and summarized in \cite{Zhou1,Zhou2}.

To describe the spin correlated BEC, 
it is most convenient to introduce 
Weyl representation of $SU(2)$ involving polynomials of a unit vector $(u,
v)$\cite{Arovas,Afflect}. 
Each unit vector is represented by a point ${\Omega}$ on a sphere
with polar coordinates $(\theta, \phi)$; namely 

\begin{eqnarray}
u=\exp(i\frac{\phi}{2})
\cos\frac{\theta}{2}, v=\exp(-i\frac{\phi}{2})\sin\frac{\theta}{2}.
\end{eqnarray}
The corresponding hyperfine spin operators are 
\begin{eqnarray}
F^+=u\frac{\partial}{\partial v}, 
F^-=v\frac{\partial}{\partial u}; \nonumber \\
F_z=\frac{1}{2}\big(u\frac{\partial}{ \partial u} -v\frac{\partial}{ 
\partial v}\big).
\end{eqnarray} 
The scalar product between two wave functions $g$ and $f$ is 
defined as $\int g^*(u,v)f(u, v)d{\Omega}/4\pi$.
(We reserve ${\bf \Omega}$ for the spin rotation discussed below.)

Under spin rotations ${\cal R}=\exp(iF_z\chi_1/2)
\exp(iF_y \theta_1/2)\\
\exp(iF_z\phi_1/2)$, $u$ and $v$ transform into

\begin{eqnarray}
&&u({\bf \Omega_1},\chi_1)=\nonumber \\
&&\exp(i\frac{\chi_1}{2})
\big(\cos\frac{\theta_1}{2} \exp(-i\frac{\phi_1}{2})u+\sin 
\frac{\theta_1}{2}\exp(i\frac{\phi_1}{2})v\big) \nonumber \\
&&v({\bf \Omega_2},\chi_2)=\nonumber \\
&&\exp(-i\frac{\chi_2}{2})
\big(-\sin\frac{\theta_2}{2}
\exp(-i\frac{\phi_2}{2})u+\cos\frac{\theta_2}{2}\exp(i\frac{\phi_2}{2})v
\big )
\nonumber \\
\end{eqnarray}
where ${\bf \Omega}_{1,2}=(\theta_{1,2}, \phi_{1,2})$.
The following identities hold
\begin{eqnarray}
&&\int u^*({\bf \Omega}_1){\bf F} u({\bf \Omega}_1) 
\frac{d{\Omega}}{4\pi}=\frac{1}{6}
{\bf \Omega}_1,
\nonumber \\
&&\int v^*({\bf \Omega}_1){\bf F} v({\bf \Omega}_1) 
\frac{d{\Omega}}{4\pi}=-\frac{1}{6}{\bf \Omega}_1.
\end{eqnarray}

Spin one wave functions are polynomials of degree two  in $u$ and $v$.
$\sqrt{3} u^2$, $\sqrt{6} uv$, $\sqrt{3} v^2$ correspond to
$m=1, 0, -1$ states. 
All $F=1$ states can also be expressed in term of $\sqrt{6}u({\bf
\Omega_1})v({\bf \Omega_2})$ with ${\bf \Omega}_{1,2}$ properly chosen.
The examples are given in Fig.1.
The comparison between the polynomial representation and the usual
vector representation is given in the Appendix A.

The Hamiltonian for spin one bosons can be written as

\begin{eqnarray}
&&{\cal H}=
-\frac{1}{2M}
\sum_\alpha
\nabla^2_\alpha 
+\sum_{\alpha,\beta}
[\frac{c_0}{2} + \frac{c_2}{2}{\bf F}_\alpha \cdot {\bf F}_\beta] 
\delta({\bf r}-{\bf r}')
\nonumber \\
&&+ \sum_\alpha {\bf F}_{z\alpha}g \mu_B H. 
\end{eqnarray}
The second term is the hyperfine spin-dependent interaction and
the last term is the coupling with an 
external magnetic field ${\bf H}=H{\bf e}_z$;
$g$ is the g-factor of bosons and $\mu_B$ is the Bohr magneton.

The wave function for N spin one Bosons can generally be written as

\begin{equation}
\Psi(\{{\bf r}_\alpha\})={\cal S}\Pi_{\alpha=1...N}
\Phi_{N_\alpha}({\bf r}_\alpha)
\sqrt{6}u_\alpha({\bf \Omega}_{1\alpha}({\bf r}_\alpha))
v_\alpha({\bf \Omega}_{2\alpha}({\bf r}_\alpha)).
\label{wavefunction}
\end{equation}  
${\cal S}$ is to symmetrize the wave function;
$N_\alpha$ labels one-particle states. The phase
$[\chi_{1}-\chi_{2}]/2$ corresponds to a gauge transformation in the complex 
field $\Phi_{N_\alpha}$ introduced above. 
Without losing generalities, we set $\chi_1=\chi_2=0$.

By taking ${\bf \Omega}_{1\alpha,2\alpha}={\bf \Omega}_{1,2}({\bf r})$ and
$\Phi_{N_\alpha}({\bf r})=\Phi({\bf r})/\sqrt{N}$ ($\Phi({\bf r})$ is a
complex scalar field), one chooses ${\bf \Omega}_{1,2},\Phi({\bf r})$ as
the collective variables of the N interacting spin one Bosons. One obtains 
a Hamiltonian for the spin and superfluid sectors

\begin{eqnarray}
&&{\cal H}={\cal H}_s +{\cal H}_{c}+{\cal H}_{ex}, 
\nonumber \\
&&{\cal H}_s=\frac{1}{2}\int d{\bf r} [\frac{1}{M}(\nabla {\bf n}({\bf r}))^2
\Phi^*\Phi + 4 c_2 {\bf L}^2({\bf r}) |\Phi^*\Phi|^2 ],
\nonumber \\
&&{\cal H}_c=\frac{1}{2}\int d{\bf r} [\frac{1}{M}|\nabla \Phi({\bf r})|^2
+ 4 c_0 |\Phi^*({\bf r})\Phi({\bf r})|^2 ],
\nonumber \\
&&{\cal H}_{ex}=g \mu_B H \int d{\bf r} 
 \Phi^*({\bf r})\Phi({\bf r}) {\bf
e}_z \cdot {\bf L}({\bf r}),
\label{Hamiltonian}
\end{eqnarray} 
Here ${\bf n}({\bf r})=({\bf \Omega}_1 + {\bf \Omega}_2)/2$,  
${\bf L}({\bf r})=({\bf \Omega}_1
-{\bf \Omega}_2)/2$.
Subscripts $s$, $c$ and $ex$ 
label spin, charge and external fields.
In the spin sector, the first term is the energy cost 
of twisting ${\bf n}$ in space, representing
the spin stiffness; the second term is the
effective 
"rotation" energy in the presence of a finite spin moment
${\bf L}$.

A local spin-phase coupling term, ${\cal H}_{sc}$, which represents
the superflow in the BECs due to the Berry's phases
under spin rotation, does exist\cite{Zhou1}.
This term, however, is linear in terms of ${\bf L}$
and quadratic in terms of spatial
gradient and is much smaller than ${\cal H}_s, {\cal H}_c$ at
the long wave length limit.
Moreover, ${\cal H}_{sc}$ vanishes in a state where ${\bf L}$ is zero.
Based on these observations, it was pointed out that the local coupling is 
negligible\cite{Zhou2}.

In the most interesting limit, we can introduce
$\Phi({\bf r})=\sqrt{\rho({\bf r})}
\exp(i\chi({\bf r}))$;
the local spin density is
${\bf l}({\bf r})={\bf L}({\bf r})\rho({\bf r})$.
${\bf n}$ and ${\bf l}$ satisfy the constraint
\begin{equation}
{\bf n}({\bf r}) \cdot {\bf  l}({\bf r})=0.
\end{equation}

$\rho$ and $\phi$,  ${\bf n}({\bf r})$ and ${\bf l}({\bf r})$ obey
the following commutation relations,

\begin{eqnarray}
&&[\rho({\bf r}),\chi({\bf r'})]=i\hbar \delta({\bf r}-{\bf r'});
\nonumber \\ \nonumber \\
&&[{\bf n}_\alpha({\bf r}), {\bf n}_\beta({\bf r}')]=0, \nonumber 
\\
&&[{\bf l}_\alpha({\bf r}), {\bf n}_\beta({\bf r}')]=i\hbar
\epsilon^{\alpha\beta\gamma}{\bf n}_\gamma \delta({\bf r}-{\bf r}'), 
\nonumber \\
&& [{\bf l}_\alpha({\bf r}), {\bf l}_\beta({\bf r}')]=i\hbar
\epsilon^{\alpha\beta\gamma}{\bf l}_\gamma \delta({\bf r}-{\bf r}'). 
\label{commutator}
\end{eqnarray}
$\epsilon^{\alpha\beta\gamma}$ is an antisymmetric tensor.
The second identity in Eq.\ref{commutator} is valid only
when ${\bf L}$ per atom is much less than unity
and ${\bf n}({\bf r})$ can be considered as a classical "vector".

The equation of motion for the superfluid and the
spin sector can be derived as

\begin{eqnarray}
&& \partial_t \chi= 4c_0 \rho -\frac{1}{M\rho}\nabla^2 \rho
+\frac{1}{2M}(\nabla \rho)^2 J,
\nonumber \\
&& \partial_t \rho= \nabla\cdot (\frac{\rho}{M} \nabla \chi ),
\nonumber \\
&& \partial_t {\bf n}=4c_2 {\bf n}\times ({\bf l}-
\frac{g\mu_B {\bf H}}{4c_2}),
\nonumber \\
&& \partial_t {\bf l}=-\frac{\rho({\bf r})}{M}
{\bf n}\times \nabla^2 {\bf n}
\nonumber \\
\label{dynamics}
\end{eqnarray}
where the low frequency and long wavelength limits have been taken. 
$J=[\chi,\rho^{-1}]/i\hbar$.
Eq.\ref{dynamics} determines the
nonlinear spin dynamics in the BECs in the long wave length limit.

The first two equations indicate 
the finite compressibility
or chemical potential
in the presence of scatterings between atoms,
and the usual
current conservation law. The first equation suggests 
the AC Josephson effects involving two condensates in
symmetry broken states.
It also implies the noncommutative nature of the phase
degree freedom $\chi$ and the Hamiltonian ${\cal H}_c$ for interacting 
atoms.
This leads to the quantum fluctuations of the condensate phase $\chi$
and symmetry retoring in a finite condensate. 
Linearizing
these equations yields excitations of the nature of
zeroth sounds, which involve a periodical compression of the density and
a periodical modulation of the phase of the condensation.
The sound velocity is $v_c=\sqrt{4c_0\rho/M}$.

The third equation shows 
that ${\bf n}$ precesses in
the presence of the spin density ${\bf l}({\bf r})$ 
due to the two-body hyperfine spin dependent scatterings.
Again, it results from the noncommutative 
nature of the spin order ${\bf n}$ and ${\cal H}_s$.
The profound phenomenon of the quantum symmetry restoring of
${\bf n}$ in a finite size spinor BEC is driven by this property.
In homogeneous BECs,
the fourth equation can be considered as the "conservation law"
for the spin density ${\bf l}({\bf r})$; the corresponding spin current
density is 
${\bf j}_\alpha=\rho M^{-1} \epsilon_{\alpha\beta\eta}
{\bf n}_\eta \nabla {\bf n}_\beta$. 
$\rho$ is the density of atoms.
For the sake of simplicity, we will assume 
the system is homogeneous from now on.

When $c_2 >0$, the "rotation" energy in ${\cal H}_{s}$ is positive. 
At the zero field limit, there exists 
a saddle point solution of Eq.\ref{Hamiltonian} for the spin sector

\begin{equation}
{\bf n}({\bf r})={\bf n}_0, 
{\bf l}({\bf r})=0.
\end{equation}
${\bf n}_0$ lies on the unit sphere.
The solution is schematically shown in Fig.1.

$u({\bf n}_0)v({\bf n}_0)$ represents state $(0, 1,0)^T$  with the 
quantization
axis pointing along ${\bf n}_0$.
It is easy to confirm that this solution at $H=0$ corresponds to the
"polar" state
found in Ref.\cite{Ho}.
By expanding Eq.\ref{dynamics} around the mean field solution,
we obtain the spin waves with a sound like spectrum $\omega=
\sqrt{4c_2\rho/M}k$, which can also be obtained in the GPE approach.
The spin wave velocity is $v_s=\sqrt{{4\rho c_2}/M}$. 

However, only the vicinity of 
point ${\bf n}_0$ on the unit sphere is practically accessible in the 
representation employed in Ref. \cite{Ho}. Eq.\ref{Hamiltonian}, on the 
other hand,
is valid for any point ${\bf \Omega}_1\sim {\bf \Omega}_2$ on the sphere.
Therefore, the effective NL$\sigma$M derived below allows us to describe
a spin correlated state far away from the one given in
the GPE approach, as promised
and provides at least a qualitative picture about
the N-body ground state wave function well beyond the 
simple mean field solution.

\vspace{-0.1cm}
\begin{figure}
\begin{center}
\epsfbox{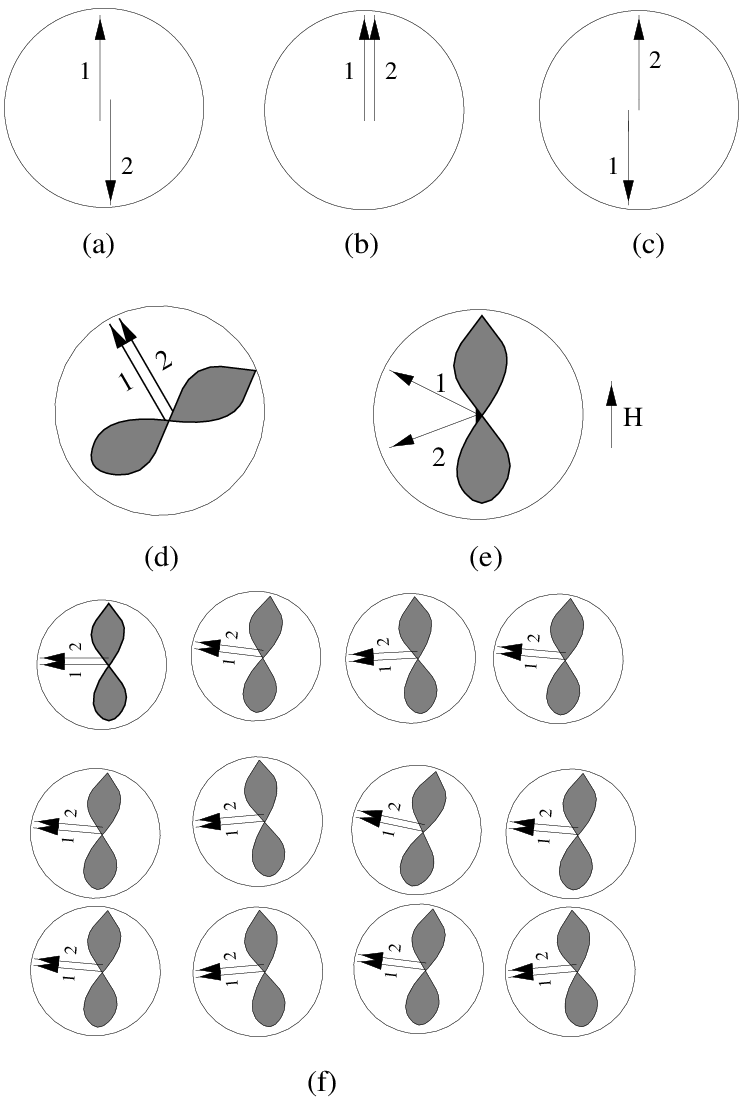}
\leavevmode
\end{center}
Fig.1 1) States represented by two unit vectors ${\bf \Omega}_1$ and ${\bf 
\Omega}_2)$:
a) $u^2$; b) $uv$; c) $v^2$.
2) ${\bf \Omega}_1$ and ${\bf \Omega}_2)$ in zero field d), and in an external
field $H$ e);
3) An example of microscopic wave functions of spin nematic states  
as indicated in f).
In a)-f),
arrows stand for ${\bf \Omega}_{1,2}$; shapes of
shaded blobs
are given by $r={3}|u({\bf \Omega})v({\bf \Omega})|^2$ in a polar 
coordinate $(r, \theta, \phi)$. 
Note in 
d) the orientation of ${\bf n}=({\bf \Omega}_1 +{\bf \Omega}_2)/2$
is arbitrary while in e) ${\bf n}$ is perpendicular to the external 
magnetic field.
\end{figure}

The low frequency sector Lagrangian for the spin and the superfluid 
component 
can be derived by taking into account Eqs.\ref{Hamiltonian},\ref{dynamics}

\begin{eqnarray}
&&{\cal L}={\cal L}_{s} +{\cal L}_c 
\nonumber \\
&&{\cal L}_c\approx \frac{\rho}{2M} [
(\nabla {\chi}({\bf r}))^2
+\frac{1}{v_c^2}(\partial_{\tau} {\chi})^2]
\nonumber \\
&&{\cal L}_s=\frac{\rho}{2M}  [
(\nabla {\bf n}({\bf r}))^2
+\frac{1}{v_s^2}(\partial_{\tau} {\bf n})^2].
\nonumber \\
\label{Lagrangian}
\end{eqnarray}
We introduce $\tau=it$ as the imaginary time. 
The non-linearlity is imposed via the constraint
$|{\bf n}^2|=1$ at a low frequency limit.
Eq.\ref{Lagrangian} is the main result of the mapping and
we keep terms which are of the lowest order in terms of $\partial_\tau$
and $\nabla$. 
${\cal L}_c$ is taken in a Gaussian
approximation and should be replaced by a
full Gross-Pitaveskii Lagrangian in general. We will be mostly 
interested in the physics in the spin sector and 
simplification in ${\cal L}_c$ doesn't affect conclusions in 
this paper.
${\cal L}_s$ in Eq.\ref{Lagrangian} represents an $o(3)$ NL$\sigma$M;
a coupling between the spin and superfluid component due to
the Berry's phase effect has been neglected\cite{Zhou1,Zhou2}.

The partition function of the Lagrangian 
can be considered as the ground state wave function,
with all the zero point fluctuations and their interactions included.
The Lagrangian ${\cal L}_s$ has been been studied extensively 
before. Many fascinating properties of the spinor BEC 
therefore can be obtained by the mapping presented in this section. 
Depending on the ratio between two-body
scattering lengths and interatomic distances, 
Eq.\ref{Lagrangian} admits solutions with very different
spin correlations. 
The symmetry broken states will be discussed in the next section and
symmetry unbroken states are to be addressed in section IV.

\subsection{$Z_2$ symmetries and $Z_2$ gauge fields}

In the previous subsection, we have shown that 
the low energy spin dynamics of the BECs of spin one atoms with 
antiferromagnetic
interactions is equivalent to that of the NL$\sigma$M. These 
BECs support
two branches spin wave excitations. In this subsection, we are going to illustrate
an extra discrete symmetry enforced by the antiferromagnetic interactions.
This discrete symmetry effectively couples
the spin and phase degrees of freedom.

Let us first look at the 
the ground state wave function shown in Eq.\ref{wavefunction} with
${\bf \Omega}_{1,2}={\bf n}$. It is 
invariant under a global transformation ${\bf n}, \chi \rightarrow 
-{\bf n}, \chi+\pi$; i.e.,
\begin{equation}
\Psi({\bf n}, \chi)=\Psi(-{\bf n}, \chi+\pi),
\Psi({\bf n})=(-1)^N \Psi(-{\bf n}), 
\label{Z2}
\end{equation}
where $\chi$ is the phase of the scalar field $\Phi(x)$ introduced in
Eq.\ref{wavefunction}.
In obtaining this symmetry, we notice $u({\bf n})=\exp(i\pi/2)
v(-{\bf n})$, with $\pi/2$ from a phase of a spin-$\frac{1}{2}$ particle
under a $180^0$ rotation.

This is certainly a quite general feature 
of all low lying excitations in $^{23}Na$ with antiferromagnetic
interactions.
It can also be illustrated using the 
zero mode Hamiltonian 

\begin{equation}
{\cal H}_{z.m.}=\rho c_2 \frac{{\bf L}^2}{2N}
\label{z.m.},
\end{equation}
${\bf L}$ is the total spin of
the BEC. And 

\begin{eqnarray}
&&[{\bf n}_\alpha, {\bf n}_\beta]=0, \nonumber 
\\&&[{\bf L}_\alpha, {\bf n}_\beta]=i\hbar
\epsilon^{\alpha\beta\gamma}{\bf n}_\gamma, 
[{\bf L}_\alpha, {\bf L}_\beta]=i\hbar
\epsilon^{\alpha\beta\gamma}{\bf L}_\gamma.
\label{nl} 
\end{eqnarray}

Eqs.\ref{z.m.},\ref{nl} show that $L$ is an angular moment operator
defined on the unit sphere spanned by ${\bf n}$.
In a spherical coordinate where ${\bf n}=(\sin\theta\cos\phi,
\sin\theta\sin\phi, \cos\theta)$, ${\bf L}$ is a differential operator 
with respect to $\theta, \phi$\cite{Zhou1}.
The eigenstates of the zero dimension Hamiltonian
are spherical harmonics of the collective variable ${\bf n}$

\begin{equation} 
\Psi({\bf n})=Y_{l,m}(\theta, \phi). 
\label{eigenstate}
\end{equation} 
The energy spectrum is given by ${\bf L}^2=l(l+1)$, which
is identical to that in \cite{Law}. The symmetry of Bosonic wave functions
imposes further constraints on parities of the states: $l=0,2,4,....N$, if
$N$ is an even number; and $l=1,3,5,...$ otherwise. Eq. \ref{eigenstate}
shows the ground state is $l=0$ or a spin singlet when $N$ is an even
number.  For an odd $N$, the ground state has a threefold degeneracy with
$l=1$. The collective behavior of the ground state depends on the
even-oddness of the number of atoms in the BEC.

\begin{figure}
\begin{center}   
\epsfbox{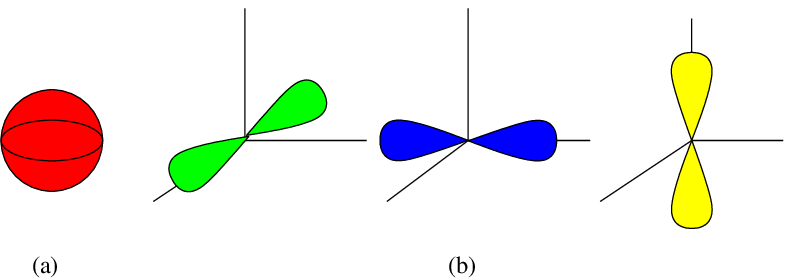}
\leavevmode
\end{center}
Fig.2 The ground states for an even a) and odd b) number of atoms in terms 
of the collective variable ${\bf n}$.
In a), the ground state is a non-degenerate $s$-orbit 
while in b), the ground state has a threefold degeneracy corresponding to
one of $p_x, p_y, p_z$ orbits, following Eq.17.
\end{figure}

Now we are ready to construct low lying wave packets, taking into account 
the symmetry of Bosonic wave functions,
\begin{eqnarray}
\Psi_N({\bf n})=\left\{ \begin{array}{cc}
\sum_{l=0,2...}A_{l,m} Y_{l,m}({\bf n}), 
& \mbox{$N$ is 
even}; \\
\sum_{l=1,3...}A_{l,m} Y_{l,m}({\bf n}), &  \mbox{$N$ is 
odd}.
\end{array} \right.
\label{parity}
\end{eqnarray}
One easily verifies that the wave function written in Eq.\ref{parity} 
observes the same $Z_2$ symmetry as a symmetry broken state.
This is not surprising from the point of view that
a symmetry broken state can be considered as a wave packet constructed out 
of the exact spectrum.

The effect of the $Z_2$ symmetry is manifestly dramatic on the dynamics
of the spinor BEC. A full description of the BEC
based on the $Z_2$ gauge fields was given in previous works,
taking into account the entanglement of the order parameter space.
So far, we haven't been able to impose 
this constraint of Eq.\ref{parity} in the function integral in a 
continuous 
limit. 
However, by introducing an optical lattice,
we are able to enforce the $Z_2$ symmetry in a mapping.
A detail derivation was presented in a preprint by
Demler et al.\cite{Zhou3}; a short account of results was given 
by Demler and Zhou\cite{Zhou4}. I will illustrate the spirit of the 
derivation 
and quote the results here.

Let me describe an optical lattice where the dynamical effects of 
the discrete symmetry can be conveniently discussed
\cite{Grynberg,Haman98,Friebel98,Rayzen97,Guidoni97,Petsas94,Deutsch98,Han00,Anderson98}.  
Experimentally, following Grynberg et al.\cite{Grynberg}, 
a body-center cubic
structure can be created by having the following four
laser beams interfere with each other: a circularly polarized light
propagating along $z$ direction and three linearly polarized beams,
with polarization plane coinciding with the $xy$ plane  propagating
in directions perpendicular to the $z$ axis.
The lattice constant in an optical lattice is determined by the wavelength 
of the 
laser.
In Gryberg et al.'s experiment, this is $514$nm; the elasticity of
the lattice is tunned by the laser intensity. 
So the model I am going to study is a realistic one from 
the experimental point of view. On the other hand,
all the phenomena
we are going to discuss should occur in both single traps
and optical lattices, though it is perhaps more practical
to observe them in a lattice setting. The feasibility was discussed  
in \cite{Zhou3}.

So we will have zero dimensional BECs with certain
number of atoms living at each site described by Eq.\ref{z.m.},
subject to the constraint in Eq.\ref{parity}.
The atoms can also hop between sites in the presence of
quantum tunneling. 
The Hamiltonian
of an array of identical optical traps is given by
\begin{eqnarray}
{\cal H} = \sum_i {\cal H}_i +\sum_{ij} {\cal H}_{ij}
\nonumber\\
{\cal H}_i =  \frac{u}{2} ( N_i - N_0 )^2
+\frac{g}{2} {\bf L}_i^2, \nonumber \\
{\cal H}_{ij} = - 2 J\, {\bf n}_i {\bf n}_j\,\, cos (\chi_i -\chi_j )
\label{lattice}
\end{eqnarray}
where 

\begin{equation}
{\bf L}_i =
i{\bf n}_i \times \frac{\partial}{\partial {\bf n}_i},
N_i=\frac{\partial}{\partial \chi_i},
\end{equation} 
and $J \approx t {N}$.
The Hamiltonian given here also includes the phase degree
of freedom to preserve the discrete symmetry under
a discrete gauge transformation.

As indicated before,
for each site characterized by the Hamiltonian in Eq.\ref{lattice}, there 
are two sorts of elementary excitations: 
a)the excitations carrying no $L$ but with $N$; b)the excitations
carrying no $N$ but with $L$. All excitations
are classified by two quantum numbers $N$ and $L$ as $(n, l)$.
The symmetry of the wave function observed by the low lying excitations
is imposed by a
constraint that the sum of $n$ and $l$ is an even integer.
So 
only $(n, 2l-n)$ are the physical excitations. 
For instance, a $(0,2)$ excitation corresponds to 
flip the spin of one of atoms without changing the number of
the atoms; a $(2,0)$ excitation is to add a singlet pair to the
condensate. And at last, a $(1,1)$ excitation represents
adding one atom of spin one to the BEC.

To carry out a functional integral in the representation of
$(n_i, l_i)$ while preserving the symmetry of the wave function,
we use a projection operator
\begin{eqnarray}
{\cal P}_i = \frac{1}{2} \sum_{\sigma_i=\pm1} 
\exp\big[ i\frac{\pi}{2}(1-\sigma_i)(n_i+l_{i})\big ],
\end{eqnarray}  
that introduces a new Ising variable $\sigma_i=\pm  1$.
The physical Hamiltonian therefore is 

\begin{equation}
{\cal H}_{phy}=
{\cal H} \prod_{i} {\cal P}_i.
\end{equation}

Projecting an enlarged space $(n, l)$
into $(n, 2m-n)$ therefore results in a set of discrete variables
which were identified as the $Z_2$ gauge fields.
The long wave length physics is characterized
by the following action,
\begin{eqnarray}
S= - \sum_{rr'} J^c_{rr'} \sigma_{rr'} cos \chi_{rr'}
   - \sum_{rr'} J^{2c}_{rr'}  cos (2 \chi_{rr'})
\nonumber\\   
   - \sum_{rr'} J^s_{rr'} \sigma_{rr'} {\bf n}_r {\bf n}_{r'}
  - \sum_{rr'} J^{2s}_{rr'}  Q^{ab}_r Q^{ab}_{r'}
\label{Sf}
\end{eqnarray}
Here
\begin{eqnarray}
Q^{ab}_r = {\bf n}^a_r {\bf n}^b_r - \frac{1}{n}\, \delta^{ab}   
\end{eqnarray}
is a nematic order parameter for the $n$-component unit vector ${\bf n}$
($n=2$  when ${\bf n}$ lies in the plane and $n=3$ when
it can rotate in all three directions). The coupling constants  
are $J^c_{r,r \pm \hat{\tau}}=(\epsilon u)^{-1}$,
$J^s_{r,r \pm \hat{\tau}}=(\epsilon g)^{-1}$,
$J^{c,s}_{r,r \pm \{\hat{x},\hat{y},\dots\}}=\epsilon J\, |\eta|$,
and  $J^{2c}_{r,r \pm \hat{\tau}}=J^{2s}_{r,r \pm \hat{\tau}}=0$,
$J^{2c,2s}_{r,r \pm \{\hat{x},\hat{y},\dots\}}=-\epsilon J /4$.
$\epsilon$ is the time unit introduced when we discretize the time;
$r=(i,\tau)$ and $r+\hat{\tau}=(i,\tau+\epsilon)$.
And $\chi_{rr'}=\chi_r-\chi_{r'}$.
Finally, $\eta$ is a parameter in the Hubbard-Stratanovich transformation
\cite{Zhou3,Zhou4}.

The site variables $\exp(i\chi)$ and ${\bf n}$ live in $S^1$ and $S^2$
respectively, representing the phase and spin degree of freedom
at each site. $\sigma_{rr'}=\pm 1 $ are discrete linking variables 
living in a $Z_2$ space. In the $S^2$ sector, we recovered the NL$\sigma$M
description of spin dynamics obtained in the previous sector.
However, under the influence of the $Z_2$ symmetry, the spin-phase degrees 
of freedom
are entangled, with interactions mediated by the $Z_2$ gauge fields.

It is obvious that the form of the action  in Eq.\ref{Sf}
is the simplest one
consistent with the charge $U(1)$, spin $SO(3)$, and gauge $Z_2$
symmetries of the model. Another term that is allowed by the $Z_2$ 
symmetry
and that is  generated by integrating out the high energy
degrees of freedom is the analogue of the Maxwell terms for the lattice 
gauge
models
\begin{eqnarray}
S_{\sigma} = -K \sum_\Box \prod_\Box \sigma_{rr'}
\label{Sg}
\end{eqnarray}
where the summation goes over plaquettes in $d+1$ dimensional lattices.
One can confirm that Eqs.\ref{Sf},\ref{Sg} are invariant under the 
following $Z_2$
gauge transformation

\begin{equation}
\chi \rightarrow \chi +\pi, {\bf n} \rightarrow -{\bf n}, 
\sigma_{rr'} \rightarrow -\sigma_{rr'}.
\label{Tran}
\end{equation}
Note that the terms proportional to $J^{2c,2s}$ are invariant under
the inversion defined in Eq.\ref{Tran}. That is, $\exp(i2\phi_r), 
Q_r^{ab}$ transform
into themselves under the $Z_2$ transformation and carry
zero $Z_2$ charges.

In the next few sections we will apply the mapping we obtained
so far to explore some aspects of the BEC of spin one atoms with
antiferromagnetic interactions.
We start with the symmetry broken states.

\section{Quantum spin nematic BECs}

The symmetry broken states of the BEC support very fascinating
topological excitations, because of the peculiar structure
of the internal space. In the next subsections, I will
explore these excitations in some detail, emphasising
on the influence of the $S^2$ and $Z_2$ symmetry in the spinor BEC.

\subsection{Topological defects in BECs}
The Lagrangian admits a symmetry broken solution 
when the two-body scattering is weak.
Topological defects are of particular interest because
of the symmetry in Eq.\ref{Z2}. They also play very important
roles in the quantum symmetry restoring and are vital for the
complete understanding of thermal and quantum phase transitions.
I will restrict myself to the static defects which
can be directly detected in an experiment.

In the previous section, two parameters, $\exp(i\chi)$ and 
${\bf n}$ are introduced for the description of BECs.
The ground state is degenerate under a gauge transformation
of the $U(1)$ group or
a rotation of ${\bf n}$ on a unit sphere $S^2$. 
Furthermore, following Eq.\ref{Z2},
the wave function is indistinguishable under a $180^0$ ${\bf n}$ rotation
and a gauge transformation: $\chi \rightarrow \chi +\pi$. 
Therefore, 
the internal space for the order parameter is
${\cal R}=[S^1\times S^2]/Z_2$; $S^1\times S^2$ factorized with respect 
to $Z_2$
is the product space of a unit circle and a unit sphere with two 
diametrically opposite
points identified. That is $({\chi}, {\bf n})=(\chi+\pi, -{\bf n})$.
This state should be defined as a quantum spin nematic state (QSNS).

The symmetry found in Eq.\ref{Z2} was not recognized in previous works
and the internal order parameter space was misidentified
as $U(1)\times S^2$ in \cite{Ho}, instead of 
$[U(1)\times S^2]/Z_2$ discussed here. 
On the other hand, for a classical nematic liquid crystal, the order 
parameter space 
is a unit sphere $S^2$ factorized with respect to $Z_2$.
The new complication in the QSNS, which originates from the
extra unit circle $S^1$ characterizing the superfluidity,
plays a very important role in topological defects.

\begin{figure}
\begin{center}
\epsfbox{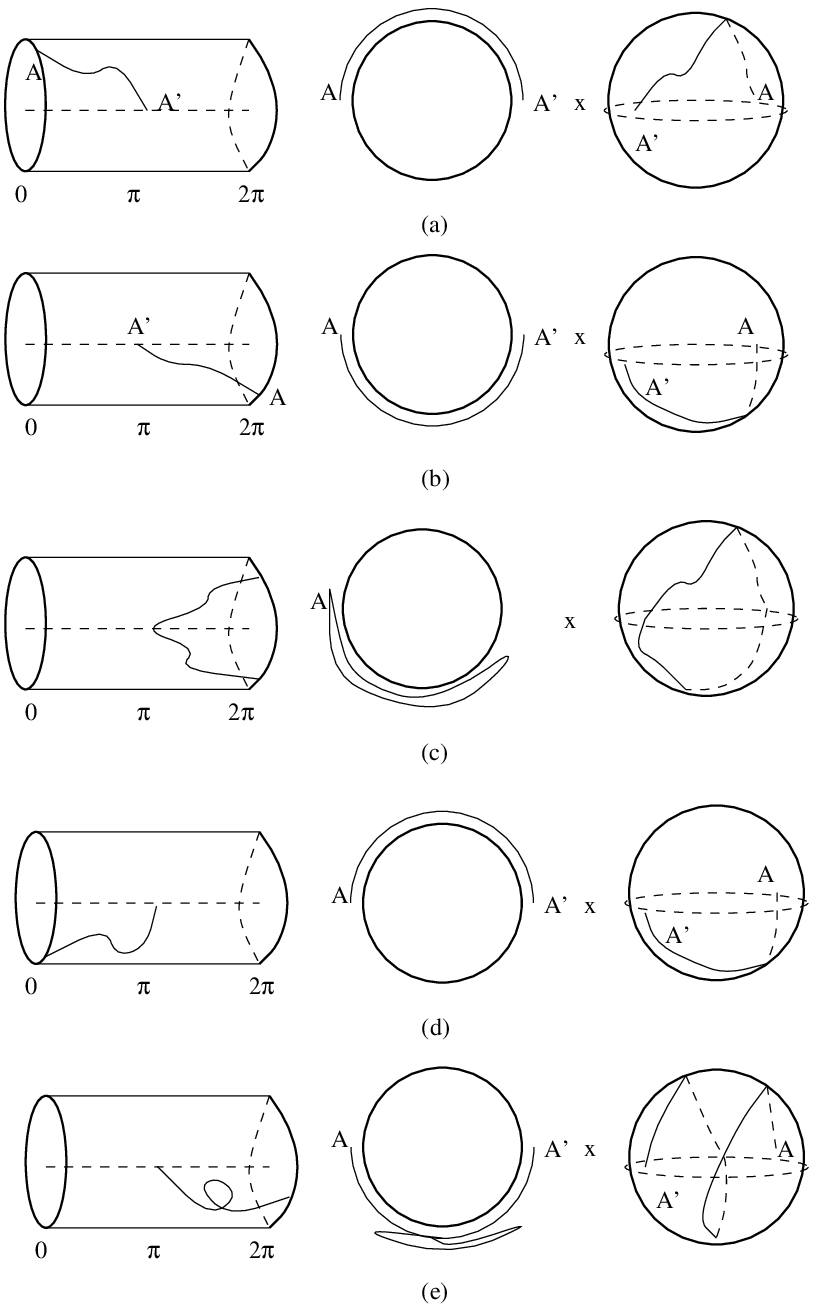}
\leavevmode
\end{center}
Fig.3. Examples of paths in ${\cal R}([S^1 \times S^2]/Z_2)$ and
their projections in 
$S^1$ and $S^2$.
$[S^1\times S^2]/Z_2$ is represented by 
a torus or a cylinder with ends identified. 
The axis of the cylinder
stands for a unit circle, along which
$\chi$ increases from 0 to $2\pi$.
Each disk, perpendicular to the axis, with boundary identified,
is homotopic to a sphere; the boundary and the center of the disk
are identified as the northern and southern pole of a two-sphere. 
Points such as $A$ and $A'$, which are diametrically opposite
in the product space, are identical.
The examples of paths homotopic to 
nonzero elements of $\pi_1([S^1\times S^2]/Z_2)$ are shown in a) 
($m=1, n=0$) (see Eq.\ref{disclination}), b) ($m=-1, n=-1$).
A path homotopic to zero element is shown in c) ($m=2, n=0$). 
Paths in d) and e) are homotopic to paths 
in a) and b) respectively. 
\end{figure}

The linear or point defects in this case are determined by 
the classes of equivalent 
maps from a circle $S^1$ or a sphere $S^2$ of a real space into 
an internal space ${\cal R}$, following the general principle for the
classification of
defects in a symmetry broken state\cite{Toulouse,Volovik,Mermin}.
The ensemble of these classes, or
the fundamental and second homotopy group
of space ${\cal R}$ are $\pi_{1,2}({\cal R})$. 
Finally, textures, which correspond to defects with no
singularities, are determined by a mapping from a three-sphere,
a three dimensional space with surface identified,
to ${\cal R}$.
Correspondingly, the spin configuration is homogeneous at infinity.
The homotopic group for textures is $\pi_3({\cal R})$.

\subsubsection{$Z_2$ strings}

The fundamental homotopy group of space $[S^1\times S^2]/Z_2$  
is a product of an integer group $Z$ and a two element group $Z_2$.
The elements of a $Z_2$ group are the indices for spin disclinations and 
$Z$ is the winding
number of superfluid vortices. A
closed path $\Gamma$ belonging to 
the nontrivial element of the $Z_2$ group is
the one connecting two diametrically opposite points 
$(\chi, {\bf n})$ and $(\chi+\pi, -{\bf n})$ in the space $[S^1\times 
S^2]$. All other paths can either be deformed
continuously into this one or into topologically trivial
ones of $Z_2$ through escaping in a third dimension(see Fig. 3).

The corresponding wave function of linear singularities
($Z_2$ strings) can be written as

\begin{eqnarray}
&&\Psi(\{\xi_\alpha \})=\Pi_\alpha \Phi(\{\xi_\alpha \})
\nonumber \\&&
(\frac{\xi_\alpha -\xi_0}{|\xi_\alpha -\xi_0|})^{n+w(m)} 
[\frac{1}{2}
Re(\frac{\xi_\alpha-\xi_0}{|\xi_\alpha-\xi_0|})^{m/2}(v^2-u^2)
\nonumber \\ &&+\frac{i}{2}
Im (\frac{\xi_\alpha-\xi_0}{|\xi_\alpha-\xi_0|})^{m/2}(v^2+u^2)]
\nonumber \\
&&\lim_{\xi \rightarrow \infty}{\bf n}(\xi)=Re 
(\frac{\xi-\xi_0}{|\xi-\xi_0|})^{m/2}
{\bf e}_{x}+ Im(\frac{\xi-\xi_0}{|\xi-\xi_0|})^{m/2}{\bf e}_{y},
\nonumber \\
&&\lim_{\xi\rightarrow \infty}{\bf v}_s(\xi)=
\frac{w(m)+n}{M|\xi-\xi_0|}
\nonumber \\&&
[Im(\frac{\xi-\xi_0}{|\xi-\xi_0|}){\bf e}_x -
Re(\frac{\xi-\xi_0}{|\xi-\xi_0|}){\bf e}_y]
\label{disclination}
\end{eqnarray}
Here $\xi=x+iy$;
$m, n$ are integers and $w(m)=1/2$ for odd $m$ and $w(m)=0$ for even $m$.
We assume lines located at 
$\xi=\xi_0=x_0+iy_0$. 
Each string is characterized in terms of $(m, n)$.
${\bf v}_{s}$ is defined as the superfluid velocity.

The $w(m)$ dependent part of 
${\bf v}_s$ is present to ensure the single valuedness
of the condensate wave function
under the spin rotation.
The spin wave function changes its sign when an atom moves
along a $Z_2$ string given in Eq.\ref{disclination},
following the identity $u({\bf n})v({\bf n})=-u(-{\bf n})v(-{\bf n})$.
The final spin wave function differs from the initial one by a
minus sign. 
In a string $(1,0)$,
${\bf v}_s$ of a half vortex is present solely to
compensate the $\pi$ phase under ${\bf n}\rightarrow -{\bf n}$
rotationally in a $Z_2$ string, as indicated in Fig.4. 
A nematic $Z_2$ string 
$(1,n)$ therefore corresponds to a $\pi$ spin disclination
superimposed by a vortex with $n+1/2$ flux quantum,
as shown in Fig.4\cite{Duncan}.

This composite structure of linear defects is uniquely associated 
with the coherence of the condensate 
and is absent in a classical nematic liquid crystal.
From an energetic point of view, 
a bare $\pi$-disclination carries a cut along which the phase
changes abruptly from $\pi$ to $2\pi$. This cut 
starting at the disclination ends only at the boundary of the coherent BEC
and costs an energy linear in terms of the size of the system. For a 
similar reason, the energy cost to separate a half vortex and
a $\pi$-disclination at distance L is linearly proportional 
to $L$. The composite strings shown in Fig.4 should be 
considered as a result of the confinement of $\pi$-disclinations and half 
vortices in the spinor BEC.

\begin{figure}
\begin{center}
\epsfbox{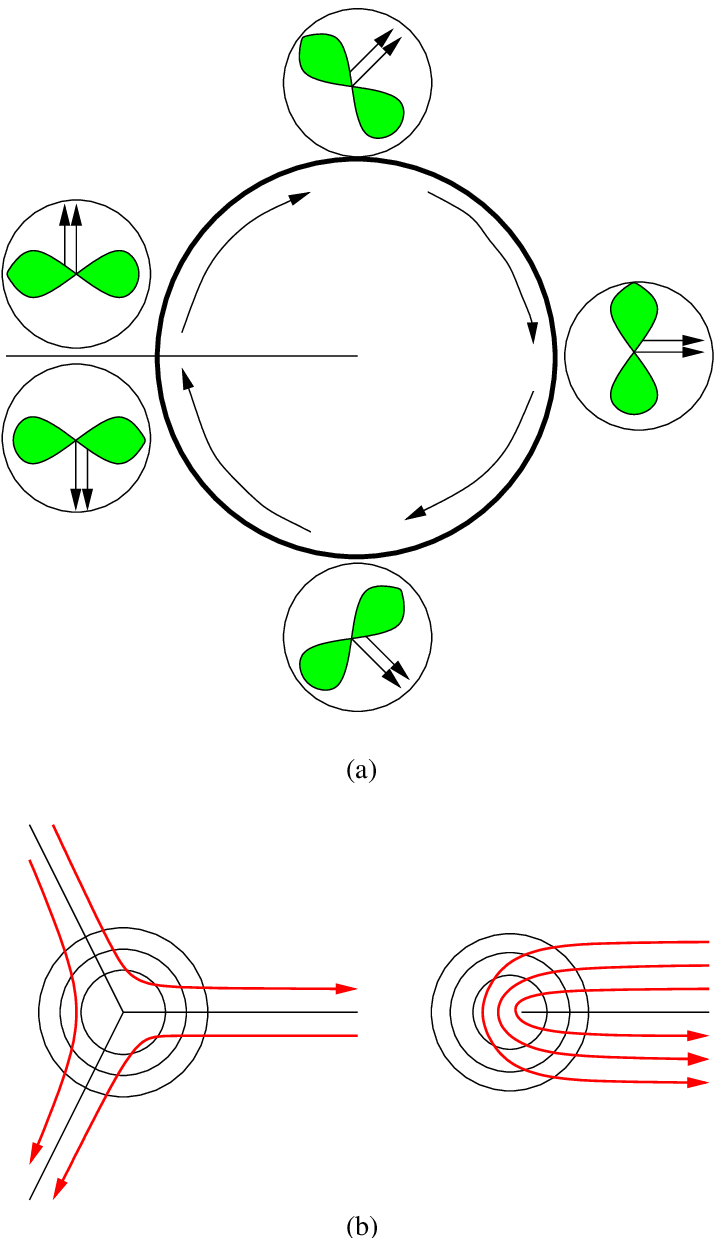}
\leavevmode
\end{center}
Fig.4  $Z_2$ strings: a) spin wave functions along a loop enclosing a 
defect (see also the caption of Fig.1); b) $\pi$ spin 
disclinations (defined in Eq.\ref{disclination}) superimposed with
superfluid vortices with half flux quanta (circular lines). 
In b), solid lines with arrows represent variation of ${\bf n}$ in defects. 
\end{figure}

The energy of a linear singularity $m=1, n=0$ is
$\rho L/4M \ln (L/a)$, with $L$ as the system size.
We should emphasis that
the spin component of the wave function carries a zero supercurrent. 
And ${\bf v}_s$ is only determined by the polynomial of degree $n+w(m)$ 
in Eq.\ref{disclination}.

$(-1, n), 
(\pm 3, n),  (\pm 5, n), ...$ linear singularities 
can be obtained from $(-1, n)$ by
continuous mappings(see examples in Fig.3).
For instance, $(-1, n)$ can be deformed into $(1, n)$ because of
"an escape in a third dimension".

For an even number $m$, ${\bf n}$ follows a closed loop on a unit 
sphere as a spin moves along the linear defects. 
The Berry's phase caused by the spin rotation along the linear
defect is zero in this case, following an
identity

\begin{equation}
Im \int_S d{\bf n}\cdot \int \frac{d\Omega}{4\pi} \frac{
\partial}{\partial {\bf n}}
u^*({\bf n})v^*({\bf n})
\times \frac{\partial}
{\partial {\bf n}}u({\bf n})v({\bf n})
=0.
\label{Berry}
\end{equation}
The integral is carried over  
the area $S$ on the unit sphere of ${\bf n}$
bounded by a path along which ${\bf n}$ varies as one moves along
the defect.

So there is no superflow of a
half integer circulation superimposed when $m$ is an even number.
And ${\bf v}_s$ vanishes in defects $(m, 0)$ 
with an even integer $m$. 
More important,
$m=\pm 2, \pm 4, ...$ can be deformed into $m=0$ configuration as shown in
Fig.3 and are homotopically identical.
There is only one family homotopically distinct
nematic spin disclination $(1, n)$.

It is particularly interesting to have a closed loop
string. The far field wave function of a closed loop linear defect 
can be expressed in a compact form as

\begin{eqnarray}
&&\Psi(\{\xi_\alpha \})=\Pi_\alpha \Phi(\{\xi_\alpha \})
u({\bf n}(\{\xi_\alpha-\xi_0\})) v({\bf n}(\{ \xi_\alpha-\xi_0 \})),
\nonumber \\
&&\lim_{|\xi| \rightarrow \infty}{\bf n}(\xi,\theta)=Re 
(\frac{\xi}{|\xi|})^{\pm 1/2}
{\bf e}_{\rho}+ Im(\frac{\xi}{|\xi|})^{\pm 1/2}{\bf e}_{z}.
\nonumber \\
&&\lim_{|\xi | \rightarrow \infty}{\bf v}_s(\xi,\theta)=0 
\label{closeloop}
\end{eqnarray}
Here ${\bf e}_{\rho}, {\bf e}_z$ are unit vectors in 
cylindrical coordinates $(z, \rho, \theta)$;
and $\xi=\rho+ i z$. The circle $\xi=\xi_0$ is the center of 
the linear defect. 
The spin rotation in a constant $\theta$ plane 
leads to a half vortex;
for a closed loop in Eq. \ref{closeloop}, the half vortex line 
forms a
closed ring along
${\bf e}_{\theta}$ direction and has no effect in the
far field.

The stability of a closed loop is a rather subtle issue
and deserves some attentions. It depends on competitions
between the energy associated with the tension of
the loop and the spin fluctuations which unfavor a 
singular structure. Under certain conditions, one indeed can show that
a close loop structure is stablized\cite{Zhou7}.

\begin{figure}
\begin{center}
\epsfbox{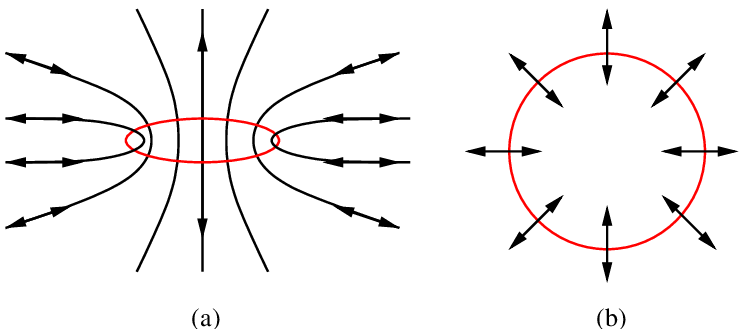}
\leavevmode
\end{center}
Fig.5 
The far fields of a closed loop $\pi$-disclination are
identical to those of a hedgehog.
The topological charge of Pontryagin type spreads 
over the entire loop while
in a hedgehog, the charge is localized at the center.
\end{figure}

However, if the loop
shrinks, one ends up with a monopole which I am turning to
in the next subsection.
The intriguing connection between a closed loop 
$\pi$ disclination and a hedgehog also lies behind a recent interesting 
work\cite{Bais}.
In the non-Abelian field theory analogy of the situation discussed here,
it was pointed out that a monopole could exist in a form 
of an Alice string.
One can examine the connection field (defined in section VI) 
of a closed loop to show
indeed the Pontryagin field strength is nonsingular at the center  
while the overall field strength at a large distance is still
of a unit charge. 
The small distance behavior is distinct from
a point-like defect.
It implies that the unit charge of the Pontryagin type is uniformly
distributed along the $\pi$ disclination. In a large loop limit, the 
linear charge density is vanishly small but with an integrated charge of 
unity. A charge exists as a closed loop string, with 
no singularities. 
Thus, the BECs of $^{23}Na$ might also be playgrounds 
for the 
physics of
Alice strings. 
A close 
analogy of the Alice string physics
was found in \cite{Zhou6} where a hypermonopole is transformed into 
an antihypermonopole when moved along a $\pi$-disclination.

\subsubsection{Hedgehogs}

The second homotopy group 
$\pi_2([S^1\times S^2]/Z_2)$ has nontrivial elements. 
Point defects are of the form of "hedgehogs" with the degree of mapping
given by $N_h$,  

\begin{eqnarray}
N_h=\frac{1}{4\pi}\int d\theta d\phi   {\bf n} \cdot \frac{\partial
{\bf n}}{\partial \theta}\times \frac{\partial {\bf n}}{\partial
\phi}. 
\end{eqnarray}

In a cylindrical coordinate $(z, \rho, \theta)$,
the far field wave function of a $N_h=1$ hedgehog 
located at the origin is 
\begin{eqnarray}
&&\Psi(\{\xi_\alpha \})=\Pi_\alpha \Phi(\{\xi_\alpha \})
[Im (\frac{\xi_\alpha}{|\xi_\alpha|}) uv +
\nonumber \\
&& \frac{1}{2}
Re \frac{\xi_\alpha}{|\xi_\alpha|}\cos(\theta_\alpha) (v^2-u^2)
+\frac{i}{2}
Re\frac{\xi_\alpha}{|\xi_\alpha|} 
\sin(\theta_\alpha) (v^2+u^2)]
\nonumber \\
&&\lim_{\xi \rightarrow \infty}{\bf n}(\xi,\theta)=Re 
\frac{\xi}{|\xi|}
{\bf e}_{\rho}+ Im \frac{\xi}{|\xi|}{\bf e}_{z},
\nonumber \\
&&\lim_{\xi \rightarrow \infty}{\bf v}_s(\xi,\theta)=0.
\end{eqnarray}
where $\xi=\rho+iz$.

However, consider a hedgehog with an index $N_h$.
One can always move ${\bf n}({\bf r})$ along a closed path $\Gamma$
in space $[S^1\times S^2]/Z_2$, which is homotopic to the nonzero element of
$\pi_1([S^1\times S^2]/Z_2)$. This connects $(\chi({\bf r}), 
{\bf n}(\bf r))$ 
into $(\chi({\bf r})+\pi, -{\bf n}({\bf r}))$ and 
transforms a hedehog with index $N_h$ into 
a $-N_h$ hedgehog.
In general, under the influence of $\pi_1([S^1 \times S^2]/Z_2)$,
the elements in $\pi_2([S^1 \times S^2]/Z_2)$ can be transformed into one
another (from $N_h$ into $-N_h$) by moving $(\chi, {\bf n})$ along the
path
$\Gamma$.
This was emphasised in the investigation of a classical nematic
liquid crystal in \cite{Volovik}, where ${\cal R}=S^2/Z_2$. 
Physically it can be done by introducing a $\pi$ spin disclination 
superimposed by a $\pi$ vortex and moving the hedgehog along 
the linear defect; thus a 
linear singularity given in Eq.\ref{disclination} defines a continuous 
deformation from
$N_h$ into $-N_h$.
The point defect is characterized only by the modulus of $N_h$
and a hedgehog with an index $N_h$ is homotopically identical to
that with $-N_h$. 
Following this discussion, the energy barrier
involved in this deformation is
the energy of the linear defect and is linear in terms of the sample size.

Despite the homotopical indistinguishability in the presence
of linear defects,
the positive and negative hedgehogs can be physically distinguished 
because of
the coherence of the BEC, unlike the situation in a classical
nematic liquid crystal. If a $\pi$ disclination is introduced 
alone, by moving a hedgehog around the linear defect, 
the condensate will acquire a $\pi$-phase with respect to the original 
one.  The $\pi$-phase difference, which can manifest itself in a Josephson
type of effect, is one of the signatures left behind by the positive or
negative hedgehogs.

The second way to distinguish the positive and negative hedgehogs is to
look at the local connection field. The idea here is
to introduce a spin-$\frac{1}{2}$ collective excitation,
which carries a half charge with respect to the connection fields.
Unlike atoms which carry zero charges, 
in a positive hedgehog configuration,
the spin-$\frac{1}{2}$
object experiences a connection field of an opposite sign
compared with that of a negative hedgehog. 
This results in the Berry's phases of different signs in the
positive and negative 
hedgehogs.

The other influence of the $Z_2$ symmetry on the point defect is that
a hedgehog is identical to 
a closed loop $Z_2$ string as illustrated before. 
Though the spin hedgehog and the closed loop $\pi$ disclination
are topologically 
identical,  
when spin fluctuations are insignificant 
the closed loop string 
has a higher energy  
because of the linear tension of a string. In this limit, we
expect a closed loop, once created, 
collapses and leaves a spin hedgehog behind.
The energy of a spin hedgehog is proportional to $\rho L/4M$,
linear in terms of the system size $L$.
Some static and dynamical aspects of the hedgehogs have also been
investigated recently in \cite{Stoof}.

The homotopical identity between a $\pi$-disclination and a hedgehog, and
the conversion between the positive and negative hedgehog in the presence
of $\pi$-disclinations are fascinating properties of the quantum spin
nematic BEC. The $Z_2$ symmetry, after all, influences every aspect of the
point defects under the consideration.

At this moment, we want to point out that
the "hedgehog" in a ferromagnetic BEC
is a "spin hedgehog" superimposed with a 
superflow. 
Along $-z$ axis, the circulation along ${\bf 
e}_\theta$ is quantized at $\int d{\bf l} \cdot \nabla \chi=1$,
representing
a vortex line ending at the monopole or a vorton discussed 
in \cite{Blaha,Volovik}.
These features are absent in the QSNS for the reason that the Berry's
phase as shown in Eq.\ref{Berry} is zero.

\subsubsection{Hopf textures}

Textures are characterized by a Hopf mapping from $S^3$ to $
[S^1\times S^2]/Z_2$ in the presence of the superfluid component. 
In this case $\pi_3([S^1 \times S^2]/Z_2) =Z$ is an integer 
group. The Hopf number which characterizes the degree of mapping 
can be written in terms of ${\bf n}$ as

\begin{eqnarray}
&& H=\frac{1}{8\pi}\int d^3{\bf r} {\bf A} \cdot \nabla\times {\bf 
A},\nonumber \\ && \nabla \times {\bf A}=
\epsilon^{abc}\epsilon_{ijk} {\bf n}_a \partial_j {\bf n}_b \partial_k 
{\bf n}_c {\bf e}_i.
\end{eqnarray}

In a cylindrical coordinate $(\rho, z, \theta)$, the
$\rho-z$ plane
is homotopical to a two-sphere,
with the northern
pole identified as the center of the texture and the southern pole as 
the boundary.
The configuration on a $\rho -z$ plane is therefore homotopically
identical to a mapping from $S^2$ to a $[S^1\times S^2]/Z_2$ 
and can be considered as a Skyrmion on the $\rho -z$ plane.
By rotating it along $z$ axis, we obtain the texture with
Hopf number $H=1$. 
The center of the texture
is located along a circle $\xi=\xi_0$ around $z$ axis
($\xi=\rho+iz$).
Similarly,
surface $Im (\xi-\xi_0)/Re (\xi-\xi_0) =const$
with boundaries identified is also homotopically identical to
a two sphere $S^2$.
Under the Hopf mapping, the 
spin configurations on these planes are the same as Skyrmions.

So in a cylindrical coordinate $(z, \rho, \theta)$,
the far field wave function for a texture with $H=1$ is 

\begin{eqnarray}
&&\Psi(\{\xi_\alpha \})=\Pi_\alpha \Phi(\{\xi_\alpha \})
\nonumber \\
&& [ \frac{1}{2}\sin g(|\xi_\alpha-\xi_0|) Re (
\frac{\xi_\alpha -\xi_0}{|\xi_\alpha -\xi_0|}\exp(i\theta_\alpha)) 
(v^2 -u^2)
\nonumber \\
&&+ \frac{i}{2}\sin g(|\xi_\alpha-\xi_0|) Im (
\frac{\xi_\alpha -\xi_0}{|\xi_\alpha -\xi_0|}\exp(i\theta_\alpha)) 
(v^2 +u^2)
\nonumber \\
&&+ \cos g(|\xi_\alpha-\xi_0|) uv ]
\nonumber \\
&&\lim_{\xi \rightarrow \infty}{\bf n}(\xi,\theta)=\sin g(|\xi-\xi_0|) [Re 
\frac{\xi-\xi_0}{|\xi-\xi_0|}{\bf e}_{\rho}+ 
Im\frac{\xi-\xi_0}{|\xi-\xi_0|}{\bf e}_{\theta}]
\nonumber \\ &&
+ \cos g(|\xi-\xi_0|){\bf 
e}_z, \nonumber\\
&&\lim_{\xi \rightarrow \infty}{\bf v}_s(\xi,\theta)=0
\end{eqnarray}
Here $g(|\xi|=0 )=0$ and $g(|\xi|=\infty)=\pi$.
Once again, the superflow ${\bf v}_s$ is zero in a Hopf texture in the QSNS
by contrast to that in a ferromagnetic BEC.

Since the order parameter is homogeneous
at the boundary of the space, the texture can considered as a "particle"
of a finite extend $\xi_0$.
The energy is estimated as $\rho \xi_0/4M$ and decreases when
the texture shrinks; this
implies an instability towards collapsing.
It can be energetically stablized only when high derivative
terms are taken into account, as suggested in \cite{Wu}.
However, this is beyond the validity of the nonlinear sigma model
where only the long wave length dynamics is correctly
characterized.

\subsection{The effects of an external magnetic field}

In the presence of an external magnetic field along $z$ direction,
${\bf \Omega}_{1,2}={\bf n}\pm {\bf e}_z g\mu_B H/4c_2 \rho$,
following Eq.\ref{dynamics};
\begin{equation}
{\bf l}=\frac{g\mu_B H}{4c_2\rho}{\bf e}_z.
\end{equation}
The Lagrangian is the same as that without an
external field (up to a constant term).
However, the constraint that ${\bf 
n}$ has to satisfy 
depends on the external field.
In the absence of an external magnetic field,
Eq.9 is automatically satisfied for ${\bf l}={\bf n}\times 
\partial_t {\bf n}/4c_2$. ${\cal L}_s$ is of the form of the
$O(3)$ NL$\sigma$M.
An external field breaks the $S^2$ symmetry and confines 
the low frequency sector of ${\bf n}$ in
a plane perpendicular to ${\bf H}$ itself, i.e.

\begin{eqnarray}
{\bf n} \cdot \frac{\mu_B H}{4c_2 \rho} {\bf e}_z=0.
\end{eqnarray}
The Lagrangian in the presence of a magnetic field
is that of an $O(2)$ NL$\sigma$M; it has the
$S^1$-symmetry at the frequency $\omega \ll \mu_B H$. 
At the high frequency limit, following the equation of motion,
${\bf n}$ precesses in a field
$4c_2 {\bf l}(x)$, much larger than the external one $H$ and the 
$S^2$-symmetry is restored.

As a consequence,
if an external magnetic field ${\bf H}=H{\bf e}_z$ is applied, 
${\bf n}_0$ is in the $xy$ plane and the parameter space for ${\bf n}_0$
is a unit circle.
The ground state is degenerate under 
the group $[U(1)\times S^1]/Z_2$, following Eq.13.
The  order parameter space for the quantum spin nematic state 
therefore is ${\cal R}=[S^1\times S^1]/Z_2$,
which represents the product space of two unit circles but
again with two diametrically opposite points $(\chi, {\bf n})$ and
$(\chi+\pi, -{\bf n})$ being identical.

The fundmental group $\pi_1([S^1 \times S^1]/Z_2)$ has
nontrivial elements.
The corresponding wave function for linear defects can be written 
in the same form as that in Eq.\ref{disclination}, $m, n$ are integers.
However in this case, all defects $(m, n)$ are 
topological distinct (at energy scales lower than the external field).  
$m=\pm 1$ and $n=\pm$ are positive and negative
$\pi$ spin disclinations, shown in Fig.2.

So far we discuss the static properties of the bulk defects.
The dynamical aspects of these defects as well as
surface defects will be addressed elsewhere.
We also notice that in the presence of the external field,
 by contrast to the classical
nematic liquid crystals, ${\bf n}$ in the QSNS
is not parallel to the external
field.
This results in sound-like collective modes even
in the presence of an external field, but with number of
modes reduced from two(at zero field) to one. 

We have restricted ourselves to the weak magnetic field limit
and neglected the possible quadratic Zeemann shift
${\cal H}_{QZ}=\sum_\alpha Q H^2 F^2_{z\alpha}$ 
(the external field $H$ is along ${\bf e}_z$ direction).
An inclusion of the quadratic Zeemann shift yields an additional 
term ${\cal L}_{QZ}=\int \rho Q H^2 ({\bf n}^2_x +{\bf n}^2_y) dx$
to the NL$\sigma$M derived in section II.
The main effect of this contribution is to align ${\bf n}$ along
the external field. Therefore, when this shift dominates,
the ground state is left with a double degeneracy:
${\bf n}={\bf e}_z$ and ${\bf n}=-{\bf e}_z$.
The spin wave develops an energy
gap of the order $QH^2$.
We will focus on the zero magnetic field case
in the rest of discussions.

\begin{figure}
\begin{center}
\epsfbox{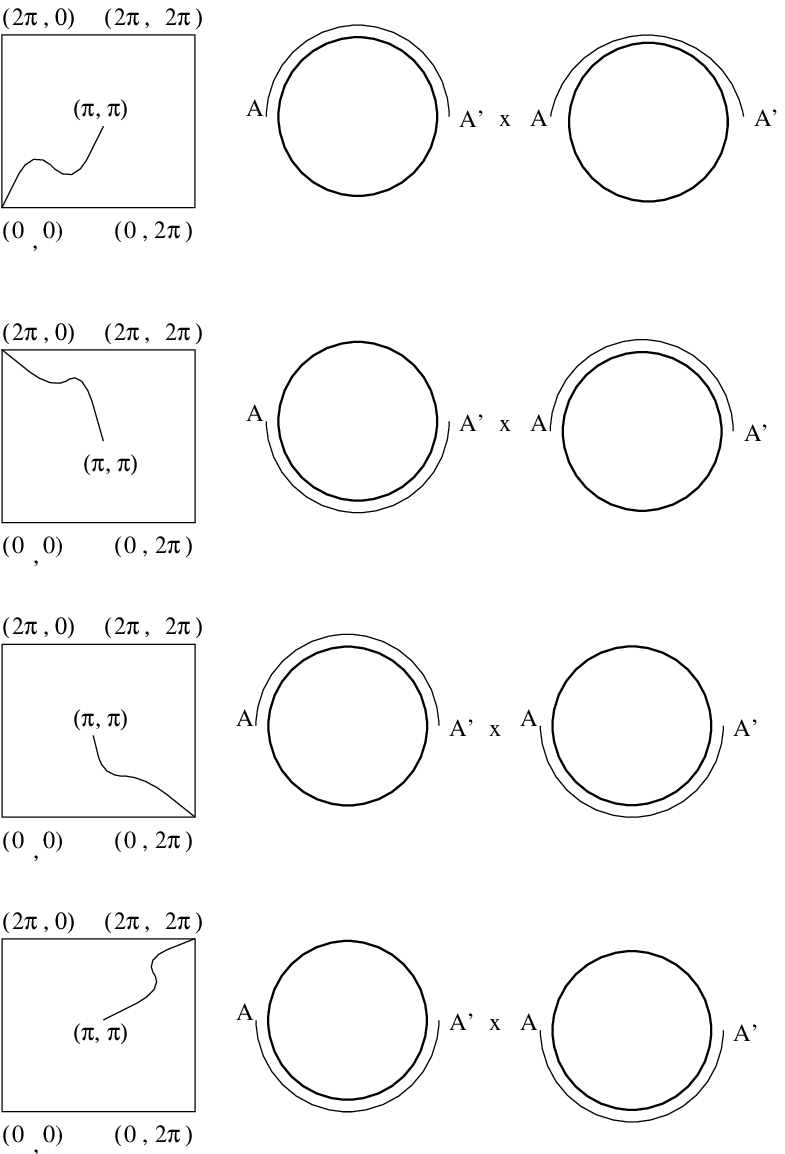}
\leavevmode
\end{center}
Fig.6. Examples of closed paths in
${\cal R}=[S^1\times S^1]/Z_2$
and projections in $S^1$ and $S^1$.
Here the surface of a torus-2, or a square with
boundaries parallel to each other identified,
stands for $[S^1\times S^1]/Z_2$.
The points such as $(\pi, \pi)$ and
$(0, 0)$, or $(0, 2\pi)$, or $(2\pi,  2\pi)$, or  $(2\pi, 0)$
are identical, representing
two diametrically opposite points in the product
space $S^1 \times S^1$. The examples of paths which are homotopic to 
nonzero
elements
of $\pi_1([S^1\times S^1]/Z_2)$ are shown in
a). (m=1, n=0), b). (m=1, n=-1), c). (m=-1, n=1), d).
(m=-1, n=0). All are topologically distinct.   
\end{figure}

\section{Spin disordered quantum condensates}

For the quantum nematic state considered in the previous sections, the
full Hamiltonian includes ${\cal H}_s$ for the spin sector ($S^2$), ${\cal
H}_c$ for the superfluid sector ($S^1$), the coupling between two sectors
${\cal H}_{sc}$ and the constraint that the wave function remains invariant
under a local $Z_2$ transformation, i.e., $\Psi(-{\bf n(x)}, \chi +
\pi)=\Psi({\bf n(x)}, \chi)$.  To study quantum nematic fluctuations or
quantum spin nematic order-disordered transitions, one has to integrate 
out
the phase degree of freedom (superfluid component) under the constraint 
and
obtain a renormalized spin dynamics.

While a full treatment of the quantum spin nematic fluctuations
taking into account the influence of the $Z_2$ symmetry 
will be present in the next section, here I will be interested 
in a spin disordered condensate
due to strong interactions between spin wave excitations in the $o(3)$ 
NL$\sigma$M;
in this limit,  
the coupling between the phase fluctuations and the spin 
fluctuations 
is irrelevant. 
The procedure is carried out self-consistently:
first we obtain results based on the o(3) NL$\sigma$M 
and then consider the influence of the $S^1$ sector and the $Z_2$ 
symmetry. 
The reduction from $[S^1\times S^2]/Z_2$ to
$S^2$ for the spin dynamics is possible based on the following 
considerations.

First,
the phase fluctuations in the BEC 
and the nematic order fluctuations
can affect each other via a direct 
coupling ${\cal L}_{sc}$.
Nevertheless, the
inclusion of ${\cal H}_{sc}$ only results in higher derivative terms;
the coupling thus becomes important only at a rather high frequency 
$\hbar \rho^{2/3}/2M$, 
which is also the upper cut-off energy of the NL$\sigma$M. 

Second, the constraint in Eq.\ref{Z2}
leads to another coupling of a pure topological origin; for instance in 
the 
static limit, it leads to superimposed 
$Z_2$ strings as discussed in section III. 
The spin nematic order-disorder transition depends
on the $Z_2$ gauge fields and spin waves in the $S^2$ sector.
However, if the phase fluctuations are weak across
the nematic order-disordered transitions driven by the zero point 
rotations of 
each atom, the $Z_2$ fields are
effectively frozen and irrelevant to the discussion.

Indeed, in $1+1$
dimension discussed in the next section, 
the nematically disordered state exists in the weakly interacting limit 
where the phase fluctuations are still negligible.
In terms of quantum tunneling, the nematic disordered state
in $1+1$ dimension is driven by texture instantons.
Quantum tunneling of the $Z_2$ strings 
(or $Z_2$ instantons) turns out to play 
little role in the nematic order-disordered transitions
because of the coherent phase sector. 
This again implies
the $Z_2$ gauge fields are not relevant to our discussion. 
In $3+1$ dimension, this is also the case and
the transitions are not driven by the $Z_2$ gauge fields if
the phases of the condensates remain coherent
around the nematic order-disorder transition point.

As in other symmetry broken states, quantum fluctuations in ${\bf n}$ 
exist in the spinor BEC because ${\bf n}$ doesn't commute with the
Hamiltonian and is not a conserved quantity.
The quantum fluctuations in the QSNS can be studied 
by considering the NL$\sigma$M,

\begin{eqnarray}
&&{\cal L}_s=\frac{1}{2f}(\partial_\mu {\bf n})^2, {\bf n} \cdot {\bf 
n}=1.
\label{NLM}
\end{eqnarray}
Here
\begin{equation}
f=(16\pi)^{1/2}(\rho
\Delta a^3)^{1/6}, \Delta a=\frac{a_2-a_0}{3}
\end{equation}
in a single trap limit. 
We also introduce dimensionless length and time:
$\tilde{{\bf r}}={\bf r}\rho^{1/3}$, $\tilde{\tau}=\tau v_s\rho^{1/3}$, and
$v_s=\sqrt{4c_2\rho/M}$.
Derivatives $\partial_\mu$ are defined as
$(\partial_{\tilde{\tau}},\partial_{\tilde x},\partial_{\tilde y},
\partial_{\tilde z})$.

In the typical low frequency fluctuations or spin waves,
$|{\bf L}|$ is much less than unity and each individual
atom remains to be in a spin state $u({\bf n}({\bf r}))v({\bf n}({\bf
r}))$,
but with ${\bf n}({\bf r}, t)$ varying in the space and time.
The quantum fluctuations of the nematic order can be
estimated in the lowest order approximation in
the weakly interacting limit. 
This is similar to the spin
wave approximation made for Heisenberg antiferromagnetic spin systems(HAFS)
\cite{Anderson52}.
The correlation function of $\delta Q^{\alpha\beta}
=Q^{\alpha\beta}-<Q^{\alpha\beta}>$ ($\alpha=x,y,z$) in the
QSNS can be calculated as

\begin{eqnarray}
&&<\delta Q^{\alpha\beta}({\bf r}, t)\delta Q^{\alpha'\beta'}(0,0)>
=\nonumber \\
&&\sum_{\eta} 
(\delta_{\alpha 0}\delta_{\alpha' 0}P_{\beta\eta} P_{\beta'\eta} 
+\delta_{0\beta}\delta_{0\beta'}
P_{\alpha\eta} P_{\alpha'\eta} 
\nonumber \\
&&+\delta_{0\alpha}\delta_{0\beta'}
P_{\beta\eta} P_{\alpha'\eta} 
+\delta_{0\beta}\delta_{0\alpha'}
P_{\alpha\eta} P_{\beta'\eta} )
{\cal C}_\eta({\bf r}, t).
\end{eqnarray}
$<>$ represents an average over all configurations.
${\cal C}_\eta
=<\delta {\bf n}_\eta({\bf r}, t)\delta {\bf n}_\eta(0,0)>$ 
reflects the dynamics of the "director"; $\eta=\pm 1$ label eigen modes
and $0$ labels the direction of ${\bf n}_0={\bf e}_z$.
$P_{\alpha\eta}={\bf e}_\alpha \cdot {\bf e}_\eta$,
where ${\bf e}_\eta$ is the unit vector of the $\eta$th eigenmode;
for ${\bf n}_0={\bf e}_z$, ${\bf e}_\eta=({\bf
e}_x+i\eta{\bf e}_y)/\sqrt{2}$.
We assume $\delta {\bf n}={\bf n}-{\bf n}_0$ is small and
different modes interact weakly.
Following Eq.\ref{NLM},

\begin{eqnarray}
&&{\cal C}_\eta({\bf r}, t)=
{f} \int d{\bf P}d\omega \frac{ \exp(i{\bf 
p}\tilde{x}-i\omega\tilde{t})}
{\omega^2 -{\bf p}^2}.
\label{fluctuation} 
\end{eqnarray}
$\omega, {\bf p}$ are dimensionless variables.
The amplitude of ${\cal C}_\eta$ is proportional to
the parameter $f$.
Eq. \ref{fluctuation} is valid when $f$ is much less than unity.
When $f$ increases, spin wave excitations start
to interact strongly and Eq.\ref{fluctuation} becomes invalid.

$f$ is a measure of the amplitude of
quantum fluctuations in the BEC of $^{23}Na$.
Following Eq.\ref{Hamiltonian}, the energy of the system consists
of two parts. a) The potential energy
$\hbar^2 \rho(\nabla {\bf n})^2 /2 M$, which is
the energy cost in the presence of a slow variation of
${\bf n}$, aligns
${\bf n}$ of different atoms. It determines the bare
spin stiffness. 
And b) the zero point
kinetic(rotation) energy
$\hbar^2 {\bf l}^2 /2 I_0$, where $I_0=(4c_2\rho/\hbar^2)^{-1}$ can be
considered as the effective
inertial of an individual atom.
This rotation energy originates from the two-body scattering
in the microscopic Hamiltonian and the inertial is
inversely proportional to the scattering length. 
The zero point rotations tend to disrupt the order of ${\bf n}$
between different atoms.
$f^{-1}$ is a square root of the ratio between the potential energy
at an interatomic scale
$\hbar^2 \rho^{2/3}/2 m$ and the zero point kinetic(rotation) energy
$\hbar^2 /2 I_0$ of an individual atom.

\subsection{Rotation symmetry restored correlated states in 3d}

At zero temperature
the $o(3)$ NL$\sigma$M in Eq.\ref{NLM} has ordered and disordered
quantum phases 
at $d > 1$,
depending on the parameter $f$.
The renormalization group (RG) equation flow is
determined by the interactions between collective
modes. The scatterings between these spin waves are determined by the 
local nonlinear spin dynamics,
especially spin waves around some slowly varying spin
configurations. 
In the following, I will take
this RG equation point of view
of the NL$\sigma$M:
$f$ is the
quantity determining spin correlations
in the ground state of the BECs.

The RG equation for $f$ can be obtained
by first integrating out fluctuations within 
$e^{-l} < |{\bf p}| < 1$, $e^{-l} <\epsilon <1$ and then 
rescaling ${\bf p}\rightarrow {\bf p} e^l$,
$\epsilon \rightarrow \epsilon e^{l}$.
In $3+1$ dimension, the RG equation takes a form

\begin{eqnarray}
\frac{ df}{dl}=-2f(f_c-f)
\end{eqnarray}
Within the framework of the NL$\sigma$M,
$f_c=8 \pi^2$ in $d=3$(see Appendix B). 
However, the exact value of $f_c$ is determined by the details of
the short-range behavior of the system and should be
obtained only by numerics.
The state where the long-range nematic order
is absent due to zero point motions of collective
variables is called a "spin disordered quantum condensate"(SDQC),
to be distinguished from the "QSNS" where the long-range order
is present.

At the strongly interacting or a high density limit $f >f_c$, the zero 
point kinetic energy dominates and the spin stiffness is renormalized 
to zero at a long wave length limit.  
Especially, in an extremely quantum disordered phase
where the scatterings between two atoms are strong and the inertial is 
small,
${\bf n}$ of each atom rotates independently and the
spin wave function of atoms only correlates at an
interatomic distance.
${\bf n}$ of each spin experiences fast diffusive
motion on the unit sphere $S^2$ with a mean free time
$1/2c_2\rho$ much shorter than $2M/\hbar^2\rho^{2/3}$, due to the strong zero
point rotation. 
The excitation spectrum has a gap
of order of $2c_2\rho$.

As the two-body scatterings get weaker,
the inertial of each individual atom $I_0$ 
becomes higher and the motion of ${\bf n}$ on
the unit sphere is slower because of a lower
zero point rotation energy. Thus,
${\bf n}$ of different atoms 
starts to correlate at a finite length $\xi$,
which is a function of the parameter $f$,
much longer
than the interatomic distance.
As this happens, the resultant effective
inertial increases proportional to the number of atoms within
the range of the correlation length, $I(\xi)\sim \xi^3/2c_2$.
And the spins of different atoms
precess a collective temporal motion on the unit sphere
with
the mean free time $\tau$. Following Eq.\ref{Hamiltonian},
${\bf L}(\xi)\sim \xi^3/c_2 \tau$ and the total rotation
energy is  ${\bf L(\xi)}^2/2I(\xi)$.
This has to be balanced by the potential energy
$\rho \xi/M$, leading
to $\tau\sim \xi/\sqrt{c_2\rho/M}$. 
It implies that in this spin correlated state,
collective excitations propagate 
with a bare spin wave velocity $\sqrt{c_2\rho/M}$ 
up to a scale $\xi$.
Finite-range nematic order exists in this case
and the correlation length can be estimated using the renormalization
approach. In the current situation, $\xi=\rho^{-1/3}f_c/[(\rho
\Delta a^3)^{1/6}-f_c]$.
Local
nematic order is established up to a distance $\xi$ in the BECs 
but {\em the state is rotationally invariant}.
A phase transition takes place at 
$\rho \Delta a^3=8\times 10^{-6} \times f_c^6$.

At weakly interacting or a low density limit $f < f_c$, the potential 
energy dominates and the spin stiffness flows to a finite value under the
renormalization group transformation.
Long-range order 
occurs. All spin one Bosons rotate as a rigid body
with an effective inertial $I(N)=N/c_2\rho$.
The energy gap in the excitation spectrum decreases accordingly

\begin{eqnarray}
E_{gap}\propto \frac{\hbar^2}{2I(N)}=\frac{\hbar^2 c_2\rho}{2N}
\end{eqnarray}
as the number of Bosons $N$ increases and vanishes as a thermal
dynamic limit is approached. 
The rotation symmetry is broken in the ground state of
the BEC as nematic long-range order is established.

We want to make the following remarks concerning
the statements made so far:

1). The quantum spin order-disorder
transition in 3d is driven by strongly interacting short wave length 
excitations. 
Therefore, in
principal, it is no longer valid to neglect the coupling between spin
and superfluidity as we did for the low frequency sector. 
We, however, believe the picture present here
is a qualitative correct one and the coupling can at most shift
the value of $f_c$. 

2). In the weakly disordered limit, the linear tension
of a $Z_2$ string at a scale $L$ is of order
$L \rho^{-1/3} E_{core}$, with the core energy $E_{core}$ 
of order $\hbar^2\rho^{2/3}/2m$. 
Large loop $Z_2$ instantons are extremely rare and can be neglected.
This justifies the reduction we made at the beginning:
The $Z_2$ gauge fields fluctuate weakly in this limit.

3) At a high density limit, one should also take into 
three-body, four-body scatterings. This can further
modify the short distance dynamics but will not
affect the conclusions arrived above in a qualitative way.

4)
One should be cautious about the definition of "phase" since
the alkali atoms under investigation are in a long lived metal stable 
gaseous state.
The life time of atomic gas is limited by three-body inelastic 
collisions\cite{Ketterle}. The collision rate increases 
dramatically as the density increases. For sodium atoms,
the life time of the metal stable gas is given as\cite{Ketterle}
$\tau_{L}=10^{-12}\times(\rho \Delta a^3)^{-2} \times 100 secs$.
If a reasonable experiment measurement can be performed at
a time scale $10ms$, then
$\rho \Delta a^3$ should be less than $10^{-4}$. 
This sets a limit on $f_c$ in order for the quantum disordered nematic
phase to be probed under the current experimental conditions. 
For $\rho \Delta a^3=10^{-6}$ as is in the experiments, 
long-range nematic order should be observed.

\subsection{Rotation symmetry retored states in 1d}

Quantum fluctuations of collective variables are most
prominent in BECs confined in highly
anisotropic traps. The N spin one bosons ground state
wave function in this limit does not live in the
zero momentum sector of the Hilbert space.
The NL$\sigma$M turns out to be the most 
powerful approach to capture spin correlations
in a rotationally invariant ground state. 
For simplicity, we approximate highly anisotropic
traps as $1d$ nematic BECs.

In $1d$ traps at zero temperature and zero field, 
one should expect there will be no long-range order 
and the state is nematically
disordered, following the renormalization group 
results of the NL$\sigma$M\cite{Polyakov}. 
The RG equation in this case can be shown 
as

\begin{eqnarray}
\frac{df}{dl}=\frac{1}{2\pi} f^2
\end{eqnarray}
which always flows into a strong coupling fixed point (disordered state)
at the low energy and long
wave length limit.
The correlation length is
$\xi=\rho^{-1}\exp(4\pi/f)$ ($\rho$ is the linear density).
That is, in a 1d trap, the excitation spectrum should
have an energy gap following \cite{Haldane83}.

In the 1d spin nematic BEC under consideration, where
the long wave length spin wave fluctuations are
most significant and ${\bf l}(x) \sim 
0$, ${\cal H}_{sc}$ can be effectively taken to be zero.
To a good approximation, the phase (or density) fluctuations and
the nematic spin fluctuations are indeed decoupled
(the $Z_2$ gauge fields be also frozen as discussed 
at the end of this section).
Therefore, the ground state can be spin disordered
but phase coherent.
The 1d BECs of $^{23}Na$ have 
long-range order in ${\chi}$ 
but short-range correlations in ${\bf n}$ 
in the weakly interacting limit.

These spin disordered states mimic the quantum spin liquid states proposed
in the literature of Heisenberg antiferromagnetic systems (HAFS). They 
represent
nematic liquids, where ${\bf n}$ of $^{23}Na$ atoms in the BEC aligns 
toward a
certain direction up to a length scale $\xi$ and the order is disrupted
afterwards due to the zero point motions originating from the spin-dependent
two-body scatterings. The zero point motions in ${\bf n}$ also lead to quantum
tunneling of spin textures, with the tunneling rate per $\xi$ inversely
proportional to the energy gap.

We notice that
the NL$\sigma$M obtained from the microscopic Hamiltonian in this paper
doesn't have a $\theta$-term
\begin{equation}
{\cal L}_\theta=\frac{\theta}{4\pi} \int d\tau dx {\bf n} \cdot 
\frac{\partial {\bf 
n}}{\partial \tau} \times \frac{\partial {\bf n}}{\partial x},
\end{equation}
which is generally present in the HAFS studied before\cite{Haldane83}.  
The presence of a $\theta$-term in the NL$\sigma$M 
can 
dramatically
affect the RG flow. 
A $\theta$-term with $\theta=\pi$, as in the half-integer HAFS,
results in an additional term in the RG equation\cite{Read}. 
In fact, the additional term
arising from a $\theta$-term  forces the RG to flow
into a massless fixed 
point in $1d$ 
and implies a power law decay of the correlation function or
gapless excitations.

The absence of a $\theta$-term in the NL$\sigma$M derived here,
which prevents the RG from flowing into a massless
fixed point and ensures an energy gap in the excitation spectrum 
of the nematic state,
can also be justified on a ground of the
$Z_2$ symmetry.
It follows a general fact that the
Berry phase in an ordered state vanishes as a result
of cancellation between $u({\bf n})$ and $v({\bf n})$
contributions, as shown in 
Eq.\ref{Berry}. 


%


We should note that
in a nematic state,
the textures with $1$ and $-1$ indices in $1+1$
dimension are homotopically indistinguishable under the influence
of the $Z_2$ symmetry.
This is similar
to what happens to the point defects
in $3d$: the spin hedgehogs with index $N_h$ and $-N_h$ are 
indistinguishable as emphasised in section IV.

The $Z_2$ symmetry in our theory also leads to
$\pi$ spin disclinations superimposed with half 
vortices in $1+1$ 
dimension: $Z_2$ 
instantons. Unlike texture instantons, the action for an individual
$Z_2$ instanton is logarithmic divergent.
When the textures are already
present, the interaction between the spin disclinations in the $Z_2$ 
instantons
becomes finite ranged due to the loss of the spin stiffness at a distance
longer than $\xi$.
However, the half vortices
in the $Z_2$ instantons still interact with each other via a logarithmic
potential when the phase remains coherent.
So the quantum tunneling of the $Z_2$ instantons
are absent in disordered nematic states in the weakly interacting limit.


The suppression of the $Z_2$ instantons
caused by the half vortices also suggests that the 
$Z_2$ gauge fields, when
coupled with a phase coherent BEC,
be "frozen" in $1+1$ dimension in the weakly interacting 
limit. The
$Z_2$ gauge field is effectively decoupled
from the spin dynamics and the spin 
order-disorder transitions are driven by the spin waves in the $S^2$ 
sector, which is self-consistent with the assumption we 
made at the beginning of the section. 
On the other hand, when both the phase and the spin sectors are 
disordered, we expect the presence of the $Z_2$ instantons in $1+1$ 
dimension.
A quantitative calculation of the multi-influence
between the $\pi$ spin disclinations and the textures, 
especially
the $\pi$ spin disclinations effect on the Haldane gap
is still absent\cite{Duncanu}.

\section{SDQCs under the influence of $Z_2$ fields}

The influence of the $Z_2$ symmetries in symmetry unbroken states,
say SDQCs, can be most conveniently examined in the lattice gauge field
model discussed in section II.
The action in Eq.\ref{Sf} represents an {\em XY} model
and a Heisenberg model coupled with a
$Z_2$ gauge field  $\sigma_{rr'}$.
In the limit when $K$ goes to infinity,
the $Z_2$ gauge fields are frozen and
$\sigma_{rr'}=1$; Eq.\ref{Sf} reduces to an {\em XY} model for the
phase sector and a Heisenberg model for the spin sector.
The spin dynamics in this case is identical to
that of an NL$\sigma$M, as discussed in details in an early  
section. The dimensionless coupling constants for the {\em XY} model and 
the NL$\sigma$M in optical lattices are
\begin{equation}
f_p^{o.l.}=\sqrt{\frac{u}{J^c}},
f_{s}^{o.l.}=\sqrt{\frac{g}{J^s}}
\end{equation}
where $g$ and $u$ as defined in Eq.\ref{lattice} depend on the spin 
susceptibility and the 
compressibility. By carrying out calculations similar to those in 
section II, one 
obtains $u=4\pi\rho (2a_2+a_0)/3M N_0$ and 
$g=4\pi\rho (a_2 - a_0)/3M N_0$ in terms of scattering lengths $a_{0,2}$,
the number density $\rho$ and the mass of atoms $M$.

In the weakly interacting limit where $f^{0.l.}_s$ is much less than 
unity,
the rotation symmetry is broken in the ground state and nematic order
occurs. 
This order leads to birefringence of a polarized
light\cite{Zhou1} and enhanced small angle light scatterings.
However, when $f^{o.l.}_s$ is much larger than unity,
the interaction between Goldstone modes
results in a spin disordered liquid. Similar to a 
resonant-valence-
bond (RVB) liquid in Heisenberg antiferromagnetic systems,
a spin disordered liquid is rotationally invariant and thus is 
free from
birefringence.

When $J^{s,c}$ approaches zero,
Eq.\ref{Sf} becomes identical to a pure
$Z_2$ gauge theory, which is dual to an Ising model\cite{Wegner}.
There are two phases for a pure $Z_2$ gauge theory at the spatial
dimension greater than one.
When $K > K_c$, the Wilson loop integral (defined in the next section)
decays exponentially as a function of an area
enclosed by the Wilson loop, i.e. in an
area law.
When $K < K_c$, the Wilson loop integral decays exponentially as the 
function of 
the perimeter of the loop.
In the former case, two $Z_2$ charges interact with a potential
linearly proportional to the distance between charges, therefore 
are confined. This has a rather profound impact on the quantum
numbers of quasiparticles which will be studied in some detail
in section VII.

In general, the BECs characterized by Eq.\ref{Sf} exhibit extremely 
rich
physics because of a coupling between the spin degree of freedom and $Z_2$
gauge fields. The influence of the $Z_2$ symmetry in symmetry broken states
was discussed in the previous section and here we want to further 
explore this influence in some symmetry partially restored states.
Particularly, we will be interested in fractionalization of
topological excitations due to the $Z_2$ gauge fields.

\subsection{Fractionalization of integer vortices}

There are three types elementary topological
defects in the BEC of $^{23}Na$\cite{Zhou3}: 

1)a half-vortex around which the phase of the
condensate changes $\pi$; 

2)a $\pi$ spin disclination around which
${\bf n}$ rotates by $180^0$;

3)a frustrated plaquette with $\prod_{\Box} \sigma_{ij}=-1$, or a
$Z_2$ vortex. A $Z_2$ vortex emits a string of $\sigma_{ij}=-1$ links
terminated either at the boundary or at another $Z_2$ vortex.

In general,
when the $U(1)$ symmetry or the $S^2$ symmetry is broken,
an individual half vortex or a $\pi$-disclination or a $Z_2$ vortex
is an energetic catastrophe because of a cut,
running from the defect to infinity, along which the phase 
of the condensate has to change abruptly.
As a result, the $Z_2$ vortices have to be confined and form bound states 
with either the half vortices or the $\pi$ disclinations in the
phases where the $U(1)$ or $S^2$ and the $Z_2$ symmetries are broken.
In a polar state studied before,
where all three symmetries are broken, only composites of the half 
vortices 
and the $\pi$ 
spin
disclinations superimposed with the $Z_2$ gauge vortices are free
objects\cite{Zhou1,Zhou3}.
On the other hand, in the states where
both the $U(1)$ and $S^2$ symmetry are unbroken,  
the $Z_2$ vortices are free excitations.

To demonstrate fractionalization of topological excitations
in the {\em SDQC} explicitly,
we consider a specific situation where the $U(1)$ symmetry is broken   
but the $S^2$ symmetry is restored.
This happens in 1d in a weakly interacting limit and also in higher
dimensions.
So the $\pi$ disclinations are always 
deconfined and are liberated.
By integrating out
${\bf n}$ of the nematic spin liquid, one obtains an effective action for
$\exp(i\chi)$,

\begin{eqnarray}
&&{\cal S}=-J_{XY}\sum_{r r'} \sigma_{r r'}   cos\chi_{r r'}  
-{K_{Z_2}}\sum_{\Box}\prod_{\Box} \sigma_{r r'}
\label{XYZ2}
\end{eqnarray}
with $J_{XY}, K_{Z_2}$ being finite, renormalized by gapped spin 
fluctuations\cite{Zhou3}.
Eq.\ref{XYZ2} represents an {\em XY} model coupled with a $Z_2$ gauge 
field.
A gauge model similar to Eq.\ref{XYZ2} was proposed for the classical
$O(3)$ nematic order-disorder phase transitions\cite{Rokhsar}.
A similar model was also considered and investigated in the context of 
cuprate high-temperature superconductors.\cite{Senthil}.

The BECs described by the reduced action in Eq.\ref{XYZ2} admit one 
ordered and two disordered phases.
At large $K_{Z_2}$ limit, a nematic order-disorder phase transition
takes place, with integer vortices condensed but 
half vortices gapped. The energy of half vortices per unit length is
finite in the disordered phase.
This state as discussed in section VII supports spin one excitations
and is a fractionalized phase.
At small $K_{Z_2}$ limit, another phase transition 
occurs, but with
half vortices condensed: the energy of a half-vortex per unit length 
vanishes
in the disordered phase. It is a confining phase where only spin two 
excitations exist.
Along the $J_{XY}=0$ axis, there is a transition between these two
disordered phases, the fractionalized one and the confining phase, as a 
pure gauge field phase transition.

When $J_{XY}$ is large, the $U(1)$ symmetry is broken.
Unlike in one-component BECs where an elementary
topological excitation carries a unit circulation,
an elementary excitation in the BEC with a hidden $Z_2$ symmetry
is a half vortex. To see this,
let us start with the limit where $K_{Z_2}$ is large but finite.
The energy density of an isolated half vortex
per unit length is 

\begin{equation}
E_{h.v.}=\frac{\rho\pi \hbar^2}{ 8M}\big( \ln \frac{L}{a} + \frac{L}{a} 
\big), 
\end{equation}
with the second term being the energy of a cut.
($L$ is the size of the system in $xy$ plane).
Moreover, two half-vortices are bound to
form an elementary excitation with a unit "flux quantum".

\begin{figure}
\begin{center}
\epsfbox{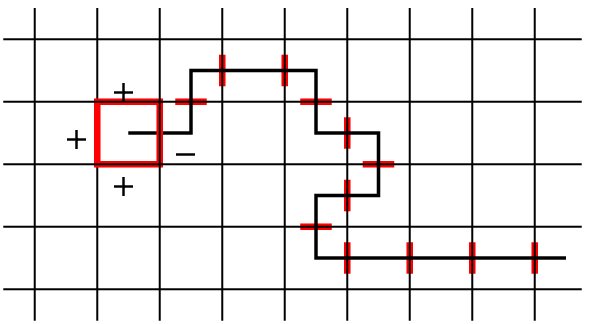}
\leavevmode
\end{center}
Fig.7 A $Z_2$ vortex emits a string of $\sigma_{ij}=-1$
links, represented by short-thick lines.
\end{figure}

However, when a $Z_2$ vortex as shown in Fig.7 is inserted,
all $\sigma_{rr'}$  along the cut  
have negative signs and
the energy catastrophe of a half vortex is removed.
The energy of a $Z_2$ vortex is from the core where 
$\prod_{\Box}\sigma_{ij}$ has a negative sign there.
The finite energy cost of the gauge vortex is proportional to 
$K_{Z_2}$(assuming $Z_2$ 
is deconfining).
The total energy of the composite in the unit of $\rho \pi\hbar^2/8M$
is just $\ln L/a$, up to a logarithmic acuracy and is one fourth of the 
energy of an integer vortex.
Therefore, an integer vortex will fractionalize into two half vortices and 
a half vortex becomes an elementary object.
Let us emphasis this arises because of the discrete $Z_2$ gauge fields and 
doesn't happen in spinless BECs or an {\em XY} model.

\subsection{BECs of singlet pairs}

Upon taking into account other
higher order processes, we have to include a hopping integral
of singlet pairs as in Eq.\ref{Sf}. 
The pair hopping integrals $J^{2s,2p}$ are particularly important when 
$g$ is large and states with  odd number
of atoms in individual wells are energetically costly:
they are not allowed  to have $L=0$ state of a rotor due to
parity constraint
and have a higher rotational energy.  It is
energetically favorable to have  even numbers of atoms on each well
arranged into singlet combinations, which can be interpreted as
binding of atoms into singlet pairs with a pair binding energy

\begin{equation}
E_p(Q)=2E(Q+1)-E(Q+2)-E(Q)= g.
\end{equation}

In the limit when $g \rightarrow \infty$
a crystal phase is possible for even numbers of atoms
per well and corresponds to a spin singlet insulator.
The superfluid phase is a condensate of singlet pairs (SSC),
in which tunneling of individual atoms between the wells is
suppressed and only singlet pairs
are delocalized.
The origin of pairing in this case is not the attraction between   
individual bosons,
but a singlet formation on the
scale of individual wells.  This reminisces
the ``attraction from repulsion'' mechanism
of electron pairing proposed by Chakravarty
and Kivelson for
high Tc cuprates, $C_{60}$, and polyacetylene \cite{Chakravarty00}.
The basic idea of this type is also the foundation of Anderson's 
interlayer tunneling mechanism for the high-$T_c$ 
cuprates\cite{Anderson92}.
In general, when a liquid is one particle-incompressible but
two-particle compressible due to either a spin gap 
as in our case, or more general due to infrared catastrophes,
the two-particle tunneling between liquids is encouraged
and long-range order of pairs can be established\cite{Anderson99}.

Formally speaking,
a large pair-tunneling term reduces 
the $U(1)$ symmetry to the $Z_2$ symmetry;
only $\phi=0$ or $\pi$ on the unit circle is 
relevant for the low energy physics. The planar spins in Eq.\ref{XYZ2} 
therefore 
have to be replaced by Ising spins $S_r=\pm 1$ and one obtains
\begin{equation}
{\cal S}=-J_{I}\sum_{rr'}\sigma_{r r'} {S}_r \cdot {S}_{r'} -
K_{Z_2}\sum_{\Box}\prod_{\Box} \sigma_{r r'}.
\label{Ising}
\end{equation}
The action in Eq.\ref{Ising} 
admits only two distinct phases, as pointed out by Fradkin and 
Shanker\cite{Fradkin}.
Along the $K_{Z_2}=\infty$ axis, a usual Ising order-disorder 
transition takes place. Along the $J_{I}=0$ axis, there will be a 
transition from
the $Z_2$ confining phase to the $Z_2$ deconfining phase.
But along the $K_{Z_2}=0$ axis, no transitions occur.
So two symmetry broken but topologically distinct phases are allowed: a) 
an Ising ordered phase with $<S_i>\neq 0$
and confining $Z_2$ 
fields and b) an Ising disordered phase with 
$<S_i>=0$
and
deconfining $Z_2$ fields.

For BECs, the first phase corresponds to 
a conventional single atom condensate 
with $<\cos\chi>\neq 0$ and the latter one is an exotic 
singlet pair condensate where $<\cos\chi>=0$ but $<\cos 
2\chi>\neq 0$, first pointed out in \cite{Zhou3,Zhou4}.
Topological excitations have distinct composition in these two symmetry 
broken states.
The Ising ordered phase breaks the $Z_2$ and $U(1)$ symmetries. The half 
vortices and the $Z_2$ gauge vortices are confined. However, in the Ising 
disordered state,  the $Z_2$ symmetry is restored while the $U(1)$ 
symmetry is 
still broken.
The half-vortex and $Z_2$ vortex are liberated and are free excitations. 
The transition between these two states is of the
Ising universality.
The other interesting aspect of the singlet pair condensate is
the a.c. Josephson effect. For the pair condensates,
\begin{equation}
\frac{d(\chi_1 -\chi_2)}{dt}=2(\mu_1 -\mu_2)
\label{Josephson}
\end{equation}
where $\chi_{1,2}$ and $\mu_{1,2}$ are the phases and chemical potentials
of two BECs. Note that the factor 2 appears in the right hand 
side
of the equation for the pair BEC.
A more realistic proposal was given in \cite{Zhou4}.

\section{Hidden topological order}

As emphasised through out this paper, the spinor BECs of spin one atoms
have many fascinating properties associated with 
the unique topology of the spin order parameter and
the entanglement of spin-phase degrees freedom. These interesting 
topological aspects result in
composited topological excitations in symmetry broken 
states, rotationally invariant spinor BECs and fractionalization of 
topological excitations in 
symmetry restored states. 

In this section, I will explore the conservation of topological 
charges.
Particularly, I want to identify hidden topological order in
spinor BECs. 
Furthermore, I generalize these ideas to the 
rotationally invariant spin liquids and show that topological  
order coexists with a short-range spin correlation, as a result
of the topological order from disorder phenomena.
Classification of electron liquids using quantum orders was recently 
reviewed by Wen\cite{Wen}. Similar ideas have been implemented for
spin triplet superconducting liquids\cite{Demler01}.

\subsection{Hidden topological order in Pontryagin fields 
in BECs}

As discussed at the beginning of the section IV, under the assumption that
the $Z_2$-strings are gapped, one can introduce a connection field
to characterize all configurations of ${\bf n}$(see also 
\cite{Zhou5,Zhou6}).
The connection field is defined as

\begin{equation}
{\bf A}({\bf r})
=\epsilon^{\alpha\beta}{\bf
e}_\alpha \nabla {\bf e}_\beta,
\end{equation}
here ${\bf e}_\alpha$ and
${\bf n}$ form a local triad; ${\bf
e}_\alpha \cdot {\bf e}_\beta=\delta_{\alpha\beta}$,
$\epsilon_{\alpha\beta} {\bf e}_\alpha {\bf e}_\beta=2{\bf n}$ and ${\bf 
e}_\alpha \cdot {\bf n}=0$.
The field strength is
a generalized gauge invariant Pontryagin density

\begin{equation}
{\bf F}_{\mu\nu}
=\epsilon^{\alpha\beta\gamma} {\bf
n}_\alpha \partial_\mu {\bf n}_\beta$$\partial_\nu {\bf n}_\gamma
\end{equation}
In $(3+1)d$, ${\bf F}_{0i}$, $i=x,y,x$ corresponds to an electric field
and $\epsilon^{ijk} {\bf F}_{ij}={\bf H}_k$
represents three components of
a magnetic
field; in $2d$, ${\bf F}_{0i}$, $i=x,y$ is an electric field with two
components and the magnetic field has only the $z$ component ${\bf 
H}_z={\bf 
F}_{xy}$.

\begin{figure}
\begin{center}
\epsfbox{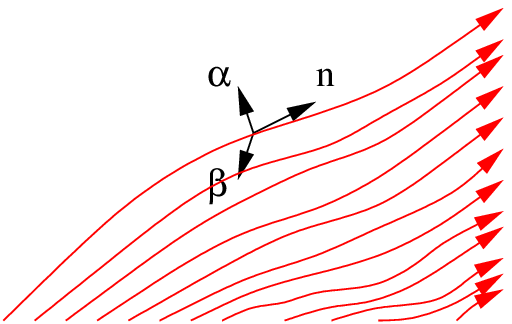}
\leavevmode
\end{center}
Fig. 8
The local triad defined at each point ${\bf r}$.
\end{figure}

One will be interested in the fluctuations of
a topological charge defined as

\begin{equation}
C_{U(1)}\big(\{ {\bf n}(\bf r)\}\big )=\frac{1}{4\pi}\oint_s dS {\bf 
H}\cdot {\bf e}_n,
\label{charge}
\end{equation}
where the integral is carried over the boundary of a large volume ${\cal 
V}$ 
and
${\bf e}_n$ is a unit vector normal to the boundary.
$C_{U(1)}$ counts the number of hedgehogs in the volume ${\cal V}$.
When the topological charge is a conserved quantity,
topological different configurations are well-defined distinguishable 
ones.
Otherwise, topological distinct configurations are converted into
each other by certain instantons.

\subsubsection{Topological order}

I am going
to show that at large ${\cal V}, t$ limit,

\begin{equation}
\lim_{t, {\cal V} \rightarrow \infty}
<\big[ C_{U(1)}\big( \{ {\bf n}({\bf r}) \}, t \big)-  
C_{U(1)}\big(\{ {\bf n}({\bf r})\}, 0\big)\big]^2>=0
\label{order}
\end{equation}
in both spin ordered and spin weakly disordered BECs, as results
of topological long-range order.
The topological order occurs when hedgehogs
are confined, either because of
the condensation of atoms or an order from
disorder mechanism. 
However, in the strongly disordered limit,
\begin{equation}
\lim_{t, {\cal V} \rightarrow \infty}
<\big[ C_{U(1)}\big(\{ {\bf n}({\bf r}) \},t\big)- 
C_{U(1)}\big(\{ {\bf n}({\bf r}) \},0 \big)\big]^2>=\frac{t{\cal 
V}}{\tau_m}
\label{disorder}
\end{equation}  
where $\tau_m$ is finite depending on the tunneling rate
of magnetic monopoles(see below).
The averages in Eqs.\ref{order},\ref{disorder} are taken over the 
many-spin 
wave function
$\Psi(\{{\bf n}\})$.
In 2d, we define 
\begin{equation}
C_{U(1)}\big(\{ {\bf n}({\bf r}) \}\big)=\frac{1}{4\pi}\int dx dy {\bf 
H}_z,
\label{charge1}
\end{equation}
the total number of Skyrmions living in the 2d space
and Eqs.\ref{order},\ref{disorder} hold in the spin
ordered
and spin disordered BECs respectively (${\cal V}$ is replaced by an area
${\cal S}$).

More formally, we can introduce an order parameter of
the Wilson loop type to characterize the hidden topological order.
The Wilson-loop integral defined as

\begin{equation}
W_{U(1)}=\big<{\cal P}\exp\big( i\oint_{\cal C} {\bf A} \cdot d{\bf r} 
\big)\big>
\end{equation}
has different asymtotical behaviors in the large loop limit in the 
presence or absence of the topological 
order. When the topological charge $C_{U(1)}$ is conserved, 
the instantons
which connect topological different configurations are forbidden;
the exponent in the Wilson loop integral is a linear function
of the perimeter of the loop.
On the other hand when $C_{U(1)}$ is not conserved,
the instantons are allowed and one can confirm the exponent 
of the Wilson loop integral is a linear function of the area enclosed by 
the loop.
Though the Wilson loop integral was originally introduced for
pure gauge fields, here we employ it for the characterization
of the topological order in the spinor BEC where gauge fields are coupled
with coherent matter fields.

The topological conservation law has 
an important consequence on collective excitation spectra
in spin disordered BECs. One can show that
when the instantons are present, the connection field has a
massive longitudinal component (with a mass $m_c$)

\begin{equation}
\big <  {\bf H}_i {\bf H}_j  \big>=(\delta_{ij}-\frac{{\bf 
k}_{i}
{\bf k}_{j}}{k^2})+\frac{{\bf k}_i {\bf k}_j}{k^2}
\frac{m_c^2}{k^2 +m_c^2}.
\end{equation}
And when the topological charge is conserved, the connection fields
have only transverse components,

\begin{equation}
\big <  {\bf H}_{i} {\bf H}_j      \big>=
(\delta_{ij}-\frac{{\bf k}_{i}{\bf k}_{j}}{{\bf 
k}^2}).
\end{equation}

The topological rigidity
can be introduced as

\begin{equation}
\rho_{t.o.}=\frac{\partial^2 E({\bf F}_{0j})}{\partial {\bf 
F}^2_{0j}}.
\label{t.o.}
\end{equation}
When topological order is present, $\rho_{t.o.}\neq 0$; otherwise,
$\rho_{t.o.}$ vanishes, indicating screening of a
small connection field (electric component).

\begin{figure}
\begin{center}   
\epsfbox{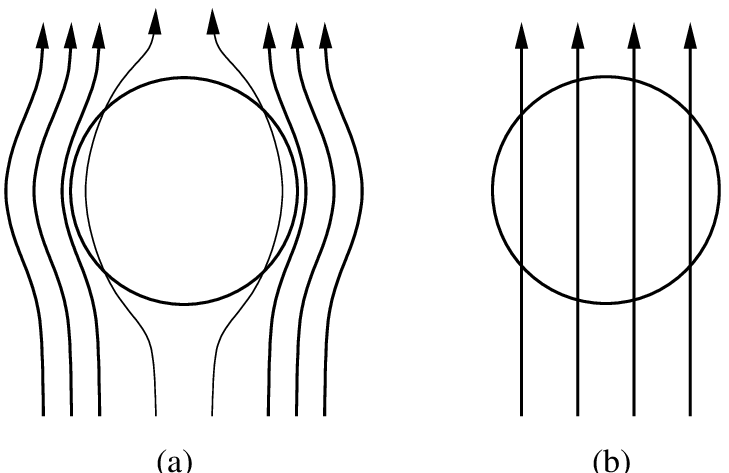}
\leavevmode
\end{center}
Fig. 9
The responses to an external connection field in the absence a) or
presence b) of topological order. 
\end{figure}

\subsubsection{Topological order and the spin stiffness}

To study the conservation of the topological charges,
we first obtain the equation of motion for ${\bf H}_k$ using 
Eqs.\ref{Hamiltonian},\ref{commutator}

\begin{equation}
\frac{\partial {\bf H}_k}{\partial t}=\epsilon_{ijk}
8 c_2 {\bf n}\cdot \partial_j {\bf n} {\bf l} \cdot \partial_i {\bf n}.
\label{Hk}
\end{equation}
Eq.\ref{Hk} demonstrates that
the topological charge is conserved if singular space-time events
are not allowed so that the product ${\bf n}\cdot \partial {\bf n}$
in the right hand side of the equation vanishes
identically.

And the topological charges fluctuate only when singular space-time 
events are permitted in the ground state.
This is rather explicit in $2d$.
First, we notice that
in space ${\bf x}=(t, {\bf r})$,
${\bf H}_{\eta}=\epsilon^{\eta\mu\nu}{\bf F}_{\mu\nu}$ is the
solution of the Gauss equation

\begin{equation}
\partial_\eta {\bf H}_{\eta}({\bf
x})=\sum_m Q_m \delta({\bf x}
-{\bf x}^m)
\label{Gauss}
\end{equation}
in the presence of space-time monopoles  $\{ {\bf x}^m \}$ 
in $(2+1)d$ Euclidean space.

The space-time monopoles represent the instantons, which essentially
connect a trivial vacuum to a Skyrmion configuration. In Euclidean space,
a space-time monopole located at $x=(t_0, {\bf r}_0)$ terminates
a Skyrmion centered at ${\bf r}_0$ at time $t_0$.
There are two important aspects in Eq.\ref{Gauss}. The first is
that the topological density is conserved in the absence of
the space-time monopoles in $(2+1)$D. Second, each monopole event causes
a change in the topological charge $C_{U(1)}$ by one unit; indeed,
following Eqs.\ref{charge1},\ref{Gauss},

\begin{equation}
\frac{\partial C_{U(1)}(\tau)}{\partial \tau}=\sum Q_m 
\delta(\tau-\tau_m),
\label{rate}
\end{equation}
where the surface contribution has been neglected since we will be
interested in the leading contribution to $C_{U(1)}$ per unit square.

\begin{figure}
\begin{center}   
\epsfbox{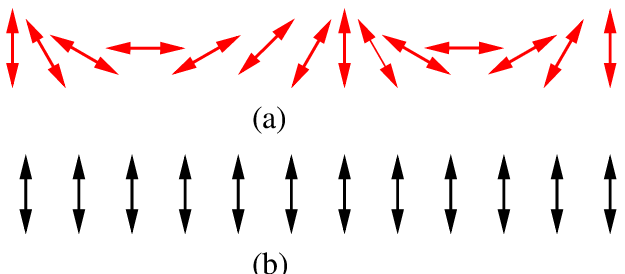}
\leavevmode
\end{center}
Fig.10  The
spin configurations 
at a) $t=+\infty$, and b) $t=-\infty$
boundaries of a space-time monopole event;
the orientation of a spin is indicated by an arrow.
\end{figure}

The probability of finding deconfined space-time  monopoles
depends on the energy of the skyrmions of $C_{U(1)}=1$ with respect to 
a $C_{U(1)}=0$
configuration.
In polar coordinates $(r, \phi_P)$,
a static Skyrmion is a configuration with 

\begin{equation}
{\bf n}(r,\phi_P)= \big
(\sin\theta(r)\cos\phi_P,
\sin\theta(r)\sin\phi_P, \cos(r)\big);
\end{equation}
$\theta(r)$ varies from $0$ at $r=0$ to $\pi$
at $r \gg \xi$, with $\xi$ an arbitrary parameter.
For spin ordered BECs where the spin Josephson effects should be 
observed, the energy of a Skyrmion is proportional to $8\pi \hbar^2 
\rho/2M$ and 
is scale invariant, as
reflected in Eq.\ref{NLM} as
well. Without lossing generality we consider a Skyrmion of a given size
$\xi$\cite{HD}.
The connection field is concentrated
in a region of the size $\xi$,

\begin{equation}
{\bf H}_z=\frac{1}{r}\sin\theta(r)
\frac{\partial \theta(r)}{\partial r},
\end{equation}
and gets screened at a length scale larger than
$\xi$. In spin ordered BECs, a Skyrmion
is nondegenerate with respect to a trivial vacuum.
As one will see, this
leads to the confinement of space-time monopoles.

To illustrate the point of the confinement,
one considers $C_{U(1)}$ as a function of time measured in some discrete 
units and
find the following binary representation for monopoles.

\begin{figure}
\begin{center}   
\epsfbox{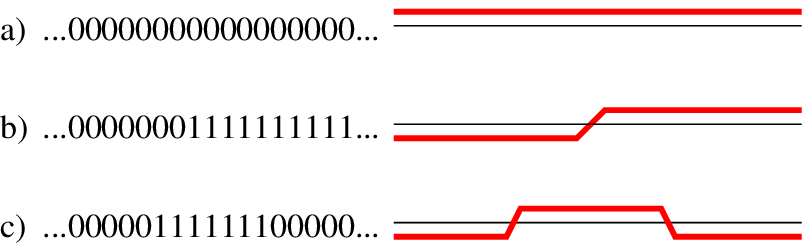}
\leavevmode
\end{center}
Fig.11 
In a binary representation,
monopole-like instantons are represented 
by kinks living on a string with $C_{U(1)}$, the topological charge
as the order parameter.
\end{figure}

Bits labeled as $0,1$ as shown in Fig.11 represent the topological charge
read out at each moment; a $1$-bit corresponds to a Skyrmion
while a $0$-bit is for a $C_{U(1)}=0$ configuration.
The a) string represents a trivial vacuum where $C_{U(1)}=0$ at all
time. In the second binary string b), a $1$-string stands for
any scale invariant Skyrmion created at the time $t_0$ defined by the
domain wall separating
the $0$-string
and $1$-string.
So a domain wall in our binary
representation stands for a monopole $Q_m=1$, terminating
a Skyrmion configuration ($C_{U(1)}=1$) at the interface and increasing 
$C_{U(1)}$ by
one unit.
A domain wall-anti domain wall pair in the third line c)
corresponds to a monopole-antimonopole pair with $Q_m=\pm 1$;
a $1$-string here, or a Skyrmion,
is terminated at a monopole at one end $t_1$ and
at an antimonopole at the other end $t_2$.

Since the energy of a Skyrmion is higher than that of a trivial
configuration, the "energy" of a $1$-bit is positive with respect to a
$0$-bit. The energy of a domain wall in Fig.11 which is the number of
$1$-bits in the structure, defines the action of the monopole event and is
proportional to the length of the $1$-string. The action of an isolated
space-time monopole event, or the creation of a Skyrmion therefore is
proportional $L_t \hbar \rho/2M$, being infinity. ($L_t$ is the perimeter
along a temporal direction.) The action of having monopole-anti-monopole
pairs is proportional to $(t_1-t_2)$, the time interval between the
creation and annihilation of the skyrmions, thus the monopoles are
confined. For this reason, the right hand side of Eq.\ref{rate} vanishes
and topological order as shown in Eq.\ref{order} 
is established.

The conclusion arrived so far can be extended to 3d straightforwardly.
The change of the topological density is caused by the quantum
nucleation of monopoles instead of Skyrmions.
In the binary representation in Fig.11, a $1$-bit stands for a
static monopole and a $0$-bit for a trivial configuration.
A domain-wall represents
termination of a static monopole at certain time.
Since the energy of a monopole is proportional
to the system size $L$ in a spin ordered BEC,
the monopole is nondegenerate
with respect to the trivial vacuum.
Following the same argument carried out in 2d,
the action to have the monopole nucleated is proportional to 
$L L_t \hbar \rho/2M$, and is
infinity.
This result has an interest impact on the monopole confinement.
The interaction between a monopole and an antimonopole
has to be linear in terms of distance between them.
The action of a monopole pair nucleated at $t_1$  and
annihilated at $t_2$ with the largest spatial separation as $|r_1-r_2|$ is
proportional to
$|r_1 -r_2|(t_1-t_2 )$ and
the monopoles are confined due to the spin stiffness.
Therefore the right hand side of Eq.\ref{rate} has to be zero because
of the confinement, and the topological charge $C_{U(1)}$ is a conserved 
quantity.
We once again arrive at Eq.\ref{order} in the spin ordered BEC.

\subsubsection{Order from disorder}

The situation in the disordered limit is more delicate and depends on
dimensionalities.
In the spin disordered BEC, the spin stiffness is renormalized
to zero and the spin
fluctuations are gapped.
However, the spin fluctuations do depend on the background 
connection fields. This is explicit in Eq. B3 in the appendix B.
Upon integrating over spin wave excitations, based on a general
consideration of the gauge invariance and the parity invariance,
we conclude the NL$\sigma$M should be reduced to

\begin{equation}
{\cal S}_s ({\bf F}_{\mu\nu})
=\frac{1}{2g_1} \int d^d{\bf r}d\tau
{\bf F}_{\mu\nu}{\bf F}_{\mu\nu} +...
\end{equation}
$g_1(\Delta_s)$ is a function of 
$\Delta_s$,  the spin gap measured in units of 
$\sqrt{c_2\rho^{5/3}/2M^2}$\cite{CP}. And in a large-N approximation,

\begin{equation}
\frac{1}{g_1} \propto \left\{ \begin{array}{cc}
\ln \Delta^{-1}_s, & \mbox{$3d$}; \\
{\Delta_s}^{-1}, & \mbox{$2d$}.
\end{array}\right.
\label{g1}
\end{equation}
We will analysis the topological order based on ${\cal S}_s$.
The energy of a $C_{U(1)}$-configuration is $\alpha(L) C_{U(1)}^2$ if
the quantum tunneling is neglected, with $\alpha(L)$ being a function of 
the system size.

In the absence of the
spin stiffness,
the energy of a Skyrmion or a monopole is determined by its interaction with 
the spin
fluctuations.
This contribution to the energy of a skyrmion
excitation in a Neel ordered antiferromagnet was previously 
considered 
by Auerbach et al.\cite{Auerbach}. The action of 
topological instantons was estimated by Murthy and Sachdev, Read
for quantum disordered antiferromagnetic spin liquids\cite{Murthy,ReadM}.

In 2d,
with the induced interactions,
the Skyrmion energy is no longer scale invariant and is minimized at 
the size $\xi=
\infty$.
The connection field of a Skyrmion spreads over the
whole 2d sheet. The energy of a Skyrmion
scales as $L^{-2}$ and
vanishes as the
system size $L$ goes to infinity.
This implies that the skyrmion configurations become degenerate with a 
trivial vacuum and 
the monopoles are deconfined.
This mechanism of deconfinement resembles the
liberation of fractionalized quasiparticles in
one dimension polymers\cite{Heeger}.
In one dimensional polyacetylene, the ground state has
a two-fold degeneracy because of the Peierls instability and
the domain wall solitons become free excitations.

Unfortunately, when the instantons are liberated and the topological charge
$C_{U(1)}$ is not conserved, the energy of
a configuration $C_{U(1)}$ is ill-defined.
A more serious consideration involves the evaluation of the action of
monopole-like instantons.
In $(2+1)d$, the partition function of a monopole
configuration $\{ {\bf x}^m \}$ 
represents point-like "charges" interacting via long-range Coulomb 
interactions.
The result suggests that the skyrmions always condense in the spin
disordered
BEC and the topological charge $C_{U(1)}$ is not a conserved quantity. It 
is important to realize that positive and negative Skyrmions can be 
physically
and homotopically distinguished because
of the coherence of the BEC, unlike the situation in a classical
nematic liquid crystal.

By moving a Skyrmion around a $\pi$-disclination,
the condensate acquires a $\pi$-phase with respect to the original
one.  The $\pi$-phase difference, which can manifest itself in a
Josephson
type of effect, is one of the signatures left behind by the positive or
negative Skyrmions.
One can also distinguish $\pm$ Skyrmions by
looking at the local connection field.
In a positive Skyrmion configuration,
a spin-$\frac{1}{2}$ collective excitation,
which carries a half charge with respect to the connection fields,
experiences a connection field of an opposite sign
compared to that of a negative Skyrmion\cite{Zhou3}.
This is similar to the situation of a point defect discussed in section 
III.

So one is able to show that the change of $C_{U(1)}$, $\partial_t{C}$, has
a short-range temporal correlation because of the instanton effects,

\begin{equation}
<\partial_t{ C}_{U(1)}\big( t \big)
\partial_t {C}_{U(1)}\big( 0 \big)>
=\exp(-\frac{t}{\tau_m}).
\label{brownian}
\end{equation}
And $\tau_m^{-1}$ is proportional to $L^2 \xi^{-2}\tau^{-1} \exp(-S_0)$.
Immediately, one recognizes that Eq.\ref{brownian} simply indicates
a random walk of $C_{U(1)}$ as a function of time
in spin disordered 2d BECs. Therefore , one concludes 
Eq.\ref{disorder} 
holds.

But in $3d$ weakly disordered BECs where the stiffness has vanished,
a static monopole
with ${\bf H}_k={\bf r}/|{\bf r}|^3$ still carries a finite energy
because of the spin fluctuations induced interactions.
In $3d$, as illustrated in ${\cal S}_s({\bf F}_{\mu\nu})$, the spin
fluctuations
discriminate
topologically different configurations. Particularly,
the fluctuations are strongest in a  $C_{U(1)}=0$ configuration
so that the energy is lowest in the topologically trivial
configuration.
The energy of a monopole
$C_{U(1)}=1$ is finite
with respect to the $C_{U(1)}=0$ configuration.

The situation differs from the 2d spin disordered limit where the energy
of a topologically nontrivial configuration with $C_{U(1)}=1$ is the 
same as
that of $C_{U(1)}=0$ one. For a monopole, each spherical shell of radius
${R}$ can be viewed as a Skyrmion squeezed into a $2d$ sheet of size
$R$. While the energy of a shell with a large radius is unsubstantial, the
smaller shells, which correspond to textures of finite sizes and have
finite energies, dominate the energy of a monopole in the spin disordered
limit. Because of the singular structure of the monopole, the coupling
between a monopole and the spin fluctuations lifts the degeneracy between
the monopole configuration and a trivial vacuum in the absence of the spin
stiffness. It also appears to us that in $3d$ the spin fluctuations should
order the topologically different configurations energetically, 
independent
of the specific form of ${\cal S}_s({\bf F}_{\mu\nu})$ we introduced. The
energy of a monopole here is inversely proportional to $g_1$, i.e. $\big(
E_o E_T \big)^{1/2} \ln \Delta^{-1}_s$. The action of having a monopole,
which leads to a change in $C_m$ by one unit, is proportional to $L_t$ and
is infinite for this reason.

\begin{figure}
\begin{center}
\epsfbox{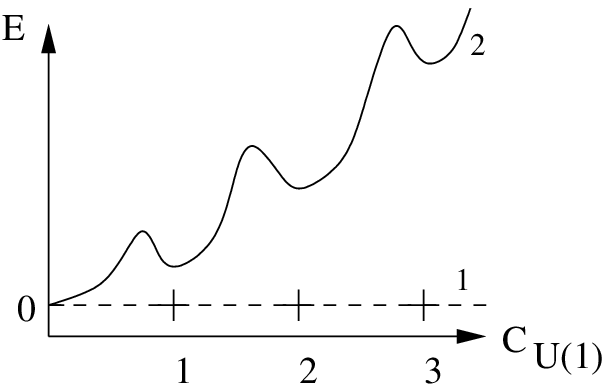}
\leavevmode
\end{center}
Fig.12 The energy of the monopoles of charge $C_{U(1)}$
in the absence of spin stiffness.  Curve 1 is the result
in a mean field approximation and curve 2 schematically represents 
the interaction energy between a monopole and spin fluctuations in a 
finite system.
\end{figure}

The monopoles remain confined even after 
the $S^2$-rotation symmetry is restored,
with the action of having
a pair of monopoles proportional to $(t_1 -t_2)$.
Only the pair production which conserves $C_{U(1)}$ is allowed.
The topological
order thus coexists with a short-range spin correlation
in $3d$.
This can be considered as a case of order from disorder
phenomena. Though topological ordering occurs in symmetry
unbroken states, it mimics the conventional order from disorder phenomena
as far as the role of the fluctuations is 
concerned\cite{Villian,Shender,Henley}.
For instance, the spectra of the spin fluctuations are affected 
by a topologically nontrivial back ground in our case and by the
symmetry breaking in \cite{Villian,Shender,Henley}.
Topological order is established as a result
of the confinement;
the confinement of the monopoles in this case is driven
purely by the spin fluctuations, instead of
the spin stiffness in spin ordered
BECs.
However, by increasing the spin disorder further, the
degeneracy between $C_{U(1)}=1$ and $C_{U(1)}=0$ could
be established. This signifies the deconfinement of the
monopoles and the breakdown of the topological order.
$C_{U(1)}$ again will have a Brownian motion as the time evolves, implied
by Eq.\ref{brownian}.

\subsection{Topological order in $Z_2$ fields}

A discussion of the hidden topological order of the $Z_2$ fields can be 
carried out parallel to the previous discussion of $U(1)$ fields. 
For simplicity, I present the result in $2d$ case, though
a generalization is possible.
It is particularly convenient to
look at the Wilson loop integral of the $Z_2$ fields, which takes a form

\begin{equation}
{\cal W}_{Z_2}=\big < \prod_{\cal C} \sigma_{rr'} \big>
\end{equation}
with the product carried out along the loop ${\cal C}$.
Once again, the exponent is a linear function of the perimeter when 
the $Z_2$ 
vortices are gapped
and is proportional to the area when the $Z_2$ vortices condense.
Consequently, the topological charge $C_{Z_2}$ in $2d$
defined as

\begin{equation}
C_{Z_2}\big( \{ \sigma_{ij}\} \big)= \prod_{\cal C} \sigma_{ij}
\end{equation}
is a conserved quantity when the $Z_2$ vortices are forbidden
dynamically and is nonconserved when the $Z_2$ vortices are allowed.

In the presence of topological order

\begin{equation}
\lim_{t, {\cal V} \rightarrow \infty}
<\big[ C_{Z_2}\big(\{\sigma_{ij}\}, t\big)-  
C_{Z_2}\big(\{\sigma_{ij}\},0 \big)\big]^2>=0,
\end{equation}

and in a topologically disordered phase

\begin{equation}
\lim_{t, {\cal V} \rightarrow \infty}
<\big[ C_{Z_2}(\{\sigma_{ij}\},t)-
C_{Z_2}(\{\sigma_{ij}\},0)\big]^2>=\frac{t{\cal V}}{\tau_m}.
\end{equation}

In BECs where either the phase or spin is ordered, the $Z_2$ vortices 
are bound with either the half-vortices or the $\pi$-spin disclinations.
The energy of the composite and the $Z_2$ vortices is
determined by the stiffness, of the phase or spin. The energy of
the $Z_2$ vortices is therefor logarithmic divergent.
In the same spirit of the previous section, we conclude that the $Z_2$ 
vortices are
forbidden and confined.
The $Z_2$ charge $C_{Z_2}$ is a conserved quantity in
the spin or phase ordered BECs, as a result of the topological 
confinement.
Following this, the Wilson loop decays in a fashion of a perimeter law.

It is perhaps more interesting to look at the possible topological order
in spin disordered BECs.
The spin fluctuations discriminate the configurations with
different $Z_2$ charges and the fluctuations are maximized in
a trivial configuration.
Therefore, the energy of a $Z_2$ charge $C_{Z_2}=1$ is higher
than a trivial one of $C_{Z_2}=0$, though the spin stiffness vanishes.

Indeed,
because of a
coupling between the $Z_2$ fields and the spin fluctuations, an effective
interaction of the $Z_2$ fields is induced upon
an integration of spin fluctuations,

\begin{equation}
{\cal S}_{in}=-\frac{1}{2g_2} 
\sum_{\Box}\prod_{\Box} \sigma_{rr'};
\label{induced}
\end{equation}
$g_2$ is a function of the spin gap developed in disordered 
limit\cite{Zhou3}.
Eq.\ref{induced} once more illustrates a general idea of the topological 
order from 
disorder.
A $Z_2$ vortex $\prod_{\Box}\sigma_{ij}$ 
does acquire an energy, by interacting with
the spin fluctuations.
The
energy of a $Z_2$ vortex in spin disordered BECs depends
on the spin gap;
it is finite but enormous in the spin weakly disordered limit in $2d$,
reflecting an energetic difficulty to have $Z_2$ vortices
due to the weakly gapped spin fluctuations. 
The rest of discussions can be carried out in a binary representation. 
The action of an $1$-string, representing termination of a $Z_2$ vortex
is infinity. The instantons connecting configurations with different $Z_2$
charges 
are suppressed and $C_{Z_2}$ is conserved.
Topological order can be established by the spin 
fluctuations.

\section{Symmetry restoring in finite size spinor BECs}

Since the experiment\cite{Stamper} was done in BECs of a few 
millions
$^{23}Na$ atoms,
it is also 
particularly interesting to consider the symmetry restoring due to a
finite size effect.
One takes a weakly interacting limit where 
quantum fluctuations of finite wave lengths are negligible.
In a zero mode approximation, 
one can neglect the spatial fluctuations of ${\bf n}$ 
and obtain from Eq.\ref{NLM}

\begin{eqnarray}
{\cal C}_\eta({\bf q}=0, t)-
{\cal C}_\eta({\bf q}=0, 0)
=-\frac{4t c_2\rho\hbar^2}{N }\int dx \frac{\sin^2 x}{x^2}
\nonumber \\
\end{eqnarray}
which is valid when the result is less than unity.
${\cal C}_\eta({\bf q}=0,0)$ should be considered of order unity.
Formally speaking, the extrapolation of the above result to $t\rightarrow
\infty$ shows that the correlator
diverges as $t$
goes to infinity. 
This implies that 
${\bf n}(t)$,  at $t \sim N /c_2\rho$(set
by the two-body scattering length),
becomes uncorrelated with the original orientation of
${\bf n}(0)$.
At longer time scales, 
${\bf n}$ precesses an 
extremely slow quantum diffusive motion on the unit sphere.
That is to say, the nematic order parameter can rotate freely in the
parameter space, as seen from ${\cal C}_\eta(0,0)$ as well
and restore the broken rotation symmetry when $N$ is finite. 

Following Eq.\ref{z.m.}, in the $0D$ BEC 
the energy gap of the low lying
excitations scales of $1/N$ at large $N$ limit.
In the ground state,
$<{\bf n}>=0$ because of the rotational invariance.
Consider
a wave packet with
${\bf n}$ confined within
a region centered at ${\bf n}_0={\bf e}_z$. The 
spread $\sqrt{<{\delta^2 \bf n}>_0}$ on 
the unit sphere at $t=0$ is very small.
The spread $<{\delta^2\bf n}>$ is defined as the expectation
value of $ {\bf n}^2_x+ {\bf n}^2_y$.
In spherical polar coordinates $(\theta,\phi)$ defined on 
a unit sphere, 
${\bf L}$ is a differential operator in terms of $(\theta, \phi)$ and
the eigenstates of ${\bf L}^2$ are spherical harmonics $Y_{\l,m}(\theta, 
\phi)$ .
The corresponding wave packet can be constructed
as

\begin{eqnarray}
&&\Psi(\theta,\phi,t)=\frac{1}{B}
\sum_{l=2n}
\exp(-\frac{l^2}{4\sigma}-i t\frac{l(l+1)c_2\rho}{2N}) 
Y_{l,0}(\theta,\phi)
\nonumber \\
&&B=\sum_l \exp(-\frac{l^2}{2\sigma}),
<\delta^2 {\bf n}>_0=\frac{A_0}{\sigma}
\label{packet}
\end{eqnarray}
where constant $A_0$ is estimated in the
Appendix C. $Y_{l,0}(\theta,\phi)$ are spheric harmonics.
Only states with even number $l$ are present in the wave packet(we
assume $N$ is an even number at the moment).
For this packet, $<{\bf n}_x>=<{\bf n}_y>=0$ and 
$<{\bf n}_z>\approx 1$ when $\sigma \gg 1$ but $\sigma/N$ is vanishingly
small.

The energy and the spread in ${\bf L}$ can be estimated 
as

\begin{equation}
\Delta E=\frac{2A_0 \sigma c_2\rho}{N},
{<\delta^2 {\bf L}>}=4A_0\sigma
\end{equation}
according to the derivation in the Appendix C.
Therefore, an excited state where ${\bf n}$ is well defined
on the unit sphere up to a spread $\sqrt{<{\delta^2 \bf n}>_0}$ has 
an infinitesimal energy cost when $N$ approaches infinity.
 
The wave packet constructed in Eq.\ref{packet}, being energetically 
extremely
close to the singlet ground state,
indicates that the singlet is very unstable with respect to
 external magnetic fields. A field of the order $o(1/N)$ 
will stablize the symmetry broken state in Eq.\ref{packet}. However,
when one deals with BEC of a small number of atoms,
the level spacing in the excitation spectrum gets bigger.
The system becomes perturbative with respective to an external 
magnetic field, as far as the disturbance is less than the energy
gap in the problem. So the claimed ultrasensitivity is
simply a consequence of nearly degenerate low lying excitations
in the large $N$ limit.
Moreover, this ultrasensitivity is absent in the
spin disordered states where a many body spin gap is opened up.

According to the
equation of motion in Eq.\ref{dynamics},  
as the time evolves, because of the two-body scatterings with $c_2>0$,
${\bf n}(t)$ gradually precesses along the finite $\delta {\bf L}$.
Indeed, the spread of ${\bf n}$ at a time $t$ becomes
\begin{equation}
<(\delta {\bf n})^2>_t=
<(\delta {\bf n})^2>_0+ {4A_0\sigma} (t\frac{c_2\rho}{N})^2,
\label{sym-res}
\end{equation}
which is valid at $t < \tau_c= \sigma /A_0\Delta E$. In an opposite limit
the spread is of order unity. 

Following Eq.\ref{sym-res}, at a time of order $1/\Delta E$,  
$\sqrt{<{\delta^2 \bf n}>_t}$ exceeds
the initial spread $\sqrt{<{\delta^2\bf n}>_0}$. 
At a longer time  $\tau_c$,
${\bf n}$ starts to explore the whole
unit sphere $S^2$ and the
rotation symmetry is restored quantum mechanically 
in the presence of spin-dependent scatterings(see Fig.13). 
A discussion is present in appendix D when an external field is applied.

\begin{figure}
\begin{center}
\epsfbox{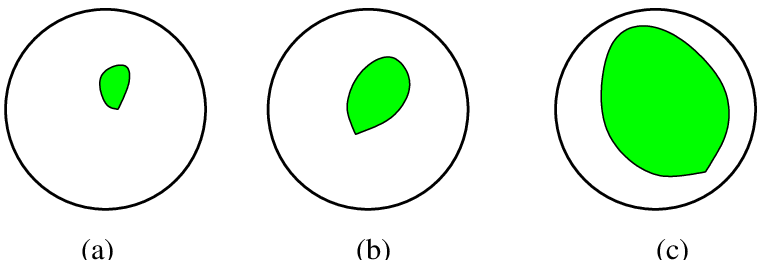}
\leavevmode
\end{center}
Fig.13 The symmetry restoring caused by the spin-dependent 
interactions($c_2>0$).
Schematically shown are the
spreads of ${\bf n}$ on a unit sphere at $t=0$ (a), $1/\Delta E$ (b),
$1/\Delta E \times 1/\sqrt{<\delta^2{\bf n}>_0}$(c).
\end{figure}

Eq.\ref{sym-res} shows how a symmetry broken state 
in finite size BECs of $^{23}Na$  becomes a
rotationally invariant state at an asymptotically long time.
For $^{23}Na$ atoms interacting via a two-body interaction with $c_2 >0$,
it was argued
in Ref.\cite{Ho} 
that in a zero mode approximation the atoms should 
condensate at $|1,m=0>$, with the rotation symmetry broken. 
On the other hand, based on an analogy to the nonlinear wave-mixing
theory, it was shown in Ref.\cite{Law} that the exact 
ground state should be a spin 
singlet with the rotation symmetry preserved. 
The symmetry restoring in a finite size system as discussed above explains
the apparent inconsistency between these previous theoretical works.

Given the number of particles as $6\times 10^6$ and $\hbar^2/I_0 \sim 
100Hz 
(500nK)$, the typical time scale for this to take place is
$t_R={NI_0}/{\hbar^2} \sim 1 \mbox{day}$.
This time scale is much longer than the life time of the trap
\cite{Ketterle}. 
Therefore the features associated with the singlet wave function 
proposed in Ref.\cite{Law}
can not be observed under that particular experiment condition.  
However, for a smaller cloud with $10^{3-4}$ atoms, the symmetry 
restoring
can take place within $10-100$secs, before the recombination takes place.
In this case, the ground state should be characterized by a singlet
wave function, instead of a saddle point solution.

Eq.\ref{sym-res} also imposes restrictions on the measurement of
a symmetry broken state.
An measurement of ${\bf n}$ in BECs of $^{23}Na$ discussed here
excites
the ground
state to an excited state where ${\bf n}$ has a finite spread on
$S^2$; the energy of this state is infinitesimal small in the
thermal dynamical limit. The time scale for ${\bf n}$ to relax,
or to restore the broken symmetry is determined by
the two-body spin dependent scattering lengths and the number
of atoms in BECs following Eq.\ref{sym-res}. Experimentally, 
the relevance of the singlet ground state
or the symmetry broken state 
depends on the life time of
the metastable gas or the interval between measurements.
Any measurement taken at a time scale longer than 
$\sigma/A_0\Delta E$
should reveal the symmetry restoring because
of zero point motions of ${\bf n}$.
Possible measurements, including measurements of birefringence 
and small angle light scatterings,
to detect orders in ${\bf n}$ in the weakly interacting
limit were discussed in \cite{Zhou1}.

It is worth emphasising that 
by contrast to a classical
nematic liquid crystal where ${\bf n}$ is pinned along the external
field and all the Goldstone modes are gapped,
${\bf n}$ can still freely rotate in a plane perpendicular to
a weak 
magnetic field in the quantum spin nematic state under consideration.
So the singularity in small angle light scattering 
survives an external magnetic field;
only the 
amplitude is reduced by a factor of one-half, when an external field is
applied. In a nematic disordered phase, the singularity
gets smeared at $q \sim \xi^{-1}$($\xi$ is the correlation
length), which can be used to study a possible phase 
transition.

\section{Discussion}

\subsection{Relation to an HAFS}

Though the NL$\sigma$M description of
the quantum spin nematic states bears some 
similarities with that of the Heisenberg antiferromagnetic spin 
system(HAFS),
there are at least two differences between the present
problem and the Heisenberg problem.
First, for the NL$\sigma$M of the HAFS,
the ground state is invariant under a rotation on a sphere $S^2$;
the order parameter space is a unit sphere $S^2$.
$\pi_1(S^2)$ is homotopic to zero and the linear singularity is
absent in this system. 
$\pi_2(S^2)=Z$ and the corresponding
point defects  are usual hedgehogs with a winding number $N$ 
distinguishable from
$-N$. 
Overall, the topological defects in the Heisenberg spin
system are qualitatively different from what we have in the QSNS 
described in section III.

However,
when a weak magnetic field is applied, 
${\bf n}$ in
the NL$\sigma$M of the HFAS lives 
on a unit circle $S^1$ and $\pi_1(S^1)=Z$ is an integer group. The linear
defects are integer vortices.
Consequently, in $2d$ in the presence of an external field, the
interaction between topological linear defects in the QSNS at an 
asymptotically
large distance decays in a same fashion as that in the HAFS. We expect
that in this specific limit, the finite temperature phase transitions are
of the same universality classes.  In addition, $\pi$-spin disclinations
superimposed with half vortices interact with each other via one half of
the interaction strength between two integer vortices. Because of the 
reduced
interaction strength between half vortices in the QSNS at $H\neq 0$, the
critical temperature is lower than that of the HAFS.

It should be noticed that
though the spin-dependent scatterings lift the degeneracy between 
different spin configurations and result in a spin nematic 
state when $c_2 >0$, the critical temperature is determined by the
density, i.e. $T_c \sim \hbar^2 
\rho^{2/3}/M$ which is independent of $c_2$ in the leading
order approximation.
When the zero point motions of ${\bf n}$ in weakly interacting
BECs are taken into account,
$T_c$ is shifted downward by an amount $o(f)$
proportional to $c_2$.

Second, in the Heisenberg antiferromagnetic system with a underlying spin
$S$, 
both the zero 
point kinetic energy ($JS$) and the potential energy ($JS^2$) depend
on the exchange integrals and the corresponding $f$, the dimensionless
coupling constant in the NL$\sigma$M,
is inversely proportional to the spin $S$ only. 
This makes observing zero temperature order-disorder phase 
transitions extremely difficult. 
In the current situation, the potential energy is independent of 
the interaction while the zero point kinetic energy is proportional
to the scattering length. 
$f$ is determined by the density and
the scattering length instead of spins of Bosons
and is a continuous variable in the QSNS. 

\subsection{Classical nematic liquids}

The classical nematic order-disordered transitions were
studied a few years ago based on a lattice gauge theory\cite{Rokhsar}.
The inversion symmetry of a nematic director is incorporated 
as a $Z_2$ gauge invariance in a lattice model,

\begin{equation} 
\frac{\cal H}{kT}=-J \sum_{rr'} \sigma_{rr'} {\bf n}_r\cdot {\bf n}_{r'}
-K\sum_{\Box} 
\prod_{\Box}{\sigma_{rr'}}
\label{Toner}
\end{equation}
where $\sigma_{rr'}=\pm 1$ are introduced as the $Z_2$ gauge fields.
The sum is again over all 
elementary plaquettes. The Hamiltonian is invariant under 
a $Z_2$ gauge transformation: ${\bf n}_r \rightarrow -{\bf n}_r$,
$\sigma_{rr'}=-\sigma_{rr'}$ (all $rr'$ links involving $r$).

The Hamiltonian in Eq. \ref{Toner}
depends on two energies scales:
$J$ as the microscopic nematic interaction and 
$K$ as the core energy of a $\pi$ spin disclination. 
At $K \rightarrow 0$, the $Z_2$ field is activated.
As $J$ decreases,  a spin nematic state is driven into
an isotropic state where the free energy of an infinity long
$\pi$ disclination is finite.
The phase transition is of the first order and believed to be the
one described by the Landau theory.
At $K\rightarrow \infty$, the $Z_2$
gauge field is frozen. In this case, as $J$ decreases, the nematic
state is driven into  another isotropic state where
the free energy of a $\pi$ disclination is proportional to the
length of it. The transition is of the second order.

The gauge theory description of the classical nematic 
liquid predicts two isotropic and optically transparent 
liquids. They are different only in topological order.
One of them has the hidden $Z_2$ order and the other one doesn't.
A recent interesting experiment by Tokyo liquid crystal
group did support this scenario\cite{Yamamoto}.
The optical turbidity measurement showed a transition between
an optically turbid phase, which apparently can be identified as
a nematic phase and an optically transparent state
which represents an isotropic liquid.
However, within the isotropic liquid, the dynamical slow down measurement 
indicates an additional phase transition between two 
optically transparent phases.
Though it is not clear at the moment whether
these two phases carry different topological order,
the situation there is quite encouraging.

\subsection{Fermionic nematic liquids}

Finally, I want to remark on other
electronic liquids which are believed to be "nematic" in nature.
The first one is the ground state of $Sr_2RuO_4$,
which was recently confirmed to be a p-wave 
superconducting state in a series of
beautiful
experiments\cite{Maeno,Mackenzie,Ishida,Duffy}.
In a p-wave superconductor of $Sr_2RuO_4$,
a Cooper pair is a spin triplet  and
experiments further
suggest that it be an Anderson-Morel
 state\cite{Vollhart}.
If a pair is considered as a composited
boson, then it occupies a
$m_F=0$ channel of spin one states, up to a rotation of the quantization 
axis.
The order parameter can be written in a tensor form as
\begin{equation}
\Delta_{\alpha\beta}=\Delta_0(T)\exp(i\chi)$ $ (\sigma_2 {\bf \sigma}
\cdot {\bf
n})_{\alpha\beta}({\bf k}_x \pm i{\bf k}_y)
\end{equation}
in the absence of spin-orbit couplings, with
${\bf \sigma}_i$, $i=0,1,2$ being Pauli matrices.
By examining this pairing wave function,
one expects
that many aspects of $Sr_2 Ru O_4$ should be very similar to
the spinor BEC described above.
In particular, the concepts of the hidden symmetries and hidden gauge
fields should be applicable in those systems\cite{Demler01}.
Of course, in addition to this, there is a new twist in p-wave
superconductors, that is the broken
time invariance and broken parity\cite{Yakovenko}.
It leads to
possible statistical transmutation of topological excitations
and quasiparticles. This remains to be understood in the context of
quantum nematic states discussed in this article.
Results on spinor BECs discussed in this article will
have no doubt to be demonstrative for the understanding of
the physics in $Sr_2RuO_4$ superconducting crystals.

The Fermionic nematic liquids
were also suggested for cuprates recently.
The nematic model discussed in \cite{Rokhsar}, and the QSNS in the spinor
BEC discussed here are intimately connected with  
the recent $Z_2$ gauge
theory developed for high-$T_c$ superconductors\cite{Senthil}. 
To understand the connection between a "fractionalized" metal
(non Fermi liquid) and a normal metal(Fermi liquid), the authors of that 
work pointed out that fractionalization
and confining issues in cuprates can be partially understood by taking 
into
account
an inversion symmetry in a spinon-chargon model.
In anisotropic phases of some strongly
correlated electron systems such as "quantum Hall nematic phases",
fermionic nematic liquids, or nematic metals were also investigated 
by Kivelson, Fradkin and their collaborators\cite{Fradkin00,Oganesyan}.
V. Organnesyan et al. recently showed the breakdown of the fermi
liquid theory in electronic nematic phases.

The investigations carried out for the spinor BEC therefore
have possible impacts on the understanding of
spin-charge separation or electron fractionalization phenomenon, 
which was first envisioned for 
cuprates about 15 years ago\cite{PWAnderson,Baskaran,Kivelson,Wiegmann}. 
As one of the most interesting inventions
by theorists in condensed matter physics, spin-charge separation
still continuously receives
a lot of attentions from condensed matter physicists.   
Though microscopically BECs studied here are very different
from those strongly correlated electronic systems,
the long wave length physics do bear many similarities to some models
employed for electron liquids.
Studies of the quantum spin nematic states,
in particular the quantum orders and spin
fractionalization\cite{Zhou5,Zhou6},
not only are 
important for the understanding of spinor BECs, but
also shed some light on some long standing
issues in many-body physics.
Being free of imperfections and easy to be manipulated magnetically or
optically, 
the QSNS might be a potentially new platform for
exploring some basic ideas in strongly correlated electron
systems, which have not been settled.

\subsection{Acknowledgement}

It is my pleasure to thank F.A. Bais, G. 't
Hooft, J. Ho, S. Kivelson, R. Moessner, M. Parikh, N. Read, 
P. Stamp, T. Senthil, J. Smit, B. Spivak, O. Starykn, D. Thouless, 
P. B. Wiegmann, B. de Wit for discussions.
I am also grateful to E. Demler and D. F. Haldane for
collaborations and many stimulating
discussions, J. Chadi and N. Wingreen for their encouragement.
This work was partially supported by ARO under DAAG 55-98-01-0270
at Princeton University and a grant from NECRI, Princeton, USA.
Finally, I want to thank 
the summer 2001 workshop of "Fundamental 
issues in quantum gases"
at ASPEN center for Physics 
and Amsterdam summer workshop 2001 on "topology and 
statistics" for their hospitalities.

\appendix

\section{Spin wave functions in $(u,v)$ representation}

In $(u,v)$ representation, states of spin $S$
are polynomials of degree $2S$ in terms of $u,v$.
Examples for $S=0,1,2$ are given below

\begin{eqnarray}
&&|S=0, S_z=0> = u^0 \rightarrow {1};
\nonumber \\
&&|S=\frac{1}{2}, S_z=\frac{1}{2}>
=\sqrt{\frac{3}{2}}u \rightarrow 
\sqrt{\frac{3}{2}}\cos\frac{\theta}{2}\exp(i\frac{\phi}{2}),   
\nonumber \\
&&|S=\frac{1}{2}, S_z=-\frac{1}{2}>
=\sqrt{\frac{3}{2}}v \rightarrow 
\sqrt{\frac{3}{2}}\sin \frac{\theta}{2}\exp(-i\frac{\phi}{2});   
\nonumber \\
\nonumber \\
&&|S=1, S_z=1>
=\sqrt{3}u^2 \rightarrow 
\sqrt{3}\cos^2\frac{\theta}{2}\exp(i{\phi}),   
\nonumber \\
&&|S={1}, S_z=0>
={\sqrt{6}}uv \rightarrow 
\sqrt{6}\sin\theta ,
\nonumber \\
&&|S=1, S_z=-1>
=-\sqrt{3}v^2 \rightarrow 
-\sqrt{3}\cos^2\frac{\theta}{2}\exp(-i{\phi}).   
\end{eqnarray}
Finally,

\begin{eqnarray}
&&\sqrt{6}u({\bf n})v({\bf n})=
\cos\theta |1, 0>+
\nonumber \\
&&\frac{1}{\sqrt{2}} \sin\theta \exp(-i\phi)|1, 1> +
\nonumber \\
&&\frac{-1}{\sqrt{2}}\sin\theta\exp(i\phi)|1,-1>
\end{eqnarray}
where ${\bf n}=(\theta, \phi)$.
At $\theta=\pi/2$, the amplitude of $|S=1,S_z=0>$ state
vanishes.

\section{Derivation of the RG equation}

Following \cite{Polyakov},
we introduce the unit vector ${\bf n}$ in a form of

\begin{eqnarray}
&&{\bf n}=(1-\sum_\eta \phi_\eta^2)^{1/2}{\bf e}_0
+\sum_\eta \phi_\eta {\bf e}_\eta,
\nonumber \\
&& {\bf e}_0 \cdot {\bf e}_\eta=0, 
{\bf e}_\eta \cdot {\bf e}_{\eta'}=\delta_{\eta\eta'}.
\nonumber \\
\end{eqnarray}
$\eta=1,2$. $\phi_\eta$ can be considered as spin waves
(fast variables) around
given slowly varying ${\bf e}_0(x)$.

Given the constraints on ${\bf e}_0, {\bf e}_\eta$ in Eq. B1,
we further introduce

\begin{equation}
\partial_\mu e_0=B_\mu^{0\eta} {\bf e}_\eta, \partial_\mu {\bf e}_\eta=
A_\mu^{\eta\eta'} {\bf e}_{\eta'} - B_\mu^{0\eta} {\bf e}_0.
\end{equation}

Inserting Eqs.B1, B2 into Eq.\ref{Lagrangian} and expanding $(1-\sum_\eta 
\phi^2_\eta)$
in terms of $\phi_\eta$, one obtains

\begin{eqnarray}
&& {\cal L}_s=\frac{1}{2f} \sum_\eta [B_\mu^{0\eta}]^2
+\frac{1}{2f}\sum_\eta(\partial_\mu \phi_\eta + 
A_\mu^{\eta\eta'}\phi_{\eta'})^2 \nonumber \\
&&+\frac{1}{2f}\sum_{\eta\eta'}
B_\mu^{0\eta}B_\mu^{0\eta'} 
[\phi_\eta \phi_{\eta'} -\delta_{\eta\eta'}\sum_\eta \phi_\eta^2].
\nonumber \\
\end{eqnarray}

By integrating out fluctuations of $\phi$ 
from $e^{-l} < |{\bf p}| <1$ and $e^{-l} < |{\epsilon}|<1$
and restoring the momentum and the energy range, we obtain
recursion relations for $f(l)$ in $d+1$ dimension

\begin{eqnarray}
&&\frac{1}{f(l)}=e^{l(d-1)}[\frac{1}{f(0)}-\Pi(l)],
\nonumber \\
&&\Pi(l)=\int_{e^{-l}}^1 \frac{d^d{{\bf q}}}{(2\pi)^d}
\frac{d\omega}{2\pi} \frac{1}{\omega^2+{\bf q}^2}.
\end{eqnarray}

The RG equation following Eq.B4 is

\begin{eqnarray}
\frac{df}{dl}=(1-d)f+{R_d}f^2,
\end{eqnarray}
with $R_1=1/2\pi, R_2=1/2\pi^2, R_3=1/4\pi^2$.

\section{Calculations of spreads at a zero field limit}
In spherical coordinates,

\begin{eqnarray}
&&{\bf L}_x=i\hbar(\sin\phi\frac{\partial}{\partial\theta}
+\cot \theta \cos\phi \frac{\partial}{\partial \phi}),
\nonumber \\
&&{\bf L}_y=i\hbar(-\cos\phi\frac{\partial}{\partial\theta}
+\cot \theta \sin\phi \frac{\partial}{\partial \phi}),
\nonumber \\
&&{\bf L}_z=i\hbar \frac{\partial}{\partial \phi}.
\nonumber \\
\end{eqnarray}

Taking into account Eq.C1, we obtain

\begin{eqnarray}
&&<(\delta^2 {\bf n})>_t=\frac{1}{B} \sum_l 
(1-M_{l,l'})\times
\nonumber \\
&&\exp(-\frac{l^2}{4\sigma}-it\frac{c_2\rho l^2}{2N})
\exp(-\frac{{l'}^2}{4\sigma}+it\frac{c_2\rho {l'}^2}{2N}),
\nonumber \\
&&M_{l',l}=\delta_{l',l}[\frac{(l+1)^2}{(2l+1)(2l+3)}
+\frac{l^2}{(2l+1)(2l-1)}]+ \nonumber \\
&&\delta_{l',l+2}[\frac{(l+1)(l+2)}{(2l+1)(2l+3)}]+
\delta_{l',l-2}[\frac{l(l-1)}{(2l+1)(2l-1)}].
\nonumber \\
\end{eqnarray}

In the limit $\sigma \gg 1$, we obtain results in Eq.\ref{packet} with
\begin{eqnarray}
&& A_0=\frac{1}{B} \sum_l  \frac{l^2}{4\sigma}\exp(-\frac{l^2}{2\sigma}).
\nonumber \\
\end{eqnarray}

\section{Calculations of $<\delta^2 {\bf n}>_t$ at $H \neq 0$}

At $H\neq 0$, the equation of motion becomes particularly 
simple. ${\bf L} \rightarrow l_\theta=i\hbar
\partial/\partial \theta$
and ${\bf n}=(\cos\theta,\sin\theta)$; 
$\theta$ is the coordinate on a unit circle.

A wave packet
of interest where $<{\bf n}_x>\sim 1$, $<{\bf n}_y>=0$ can be written as
\begin{eqnarray}
\Psi({\bf n}, t)=\frac{1}{B} \sum_{l=0,
\pm2...}\exp(-\frac{l^2}{4\sigma})
\exp(i l \theta -i t l^2 \frac{c_2\rho }{2N})\nonumber \\
\end{eqnarray}
where $\sigma \gg 1$ is the spread of $l_\theta$ in the wave packet.
One can again show that the energy, spin and 
the corresponding spread of ${\bf n}$ are

\begin{eqnarray}
&&\Delta E=\frac{2\sigma c_2\rho}{N}, 
<l_\theta^2>=4\sigma A_0 ;\nonumber \\
&&<\delta^2{\bf n}_y>_t=
A_0[\frac{1}{\sigma}+4\sigma (\frac{tc_2\rho}{N})^2].
\nonumber \\
\end{eqnarray}
$A_0(\sigma)$ is a quantity of unity.
The last equation in Eq.D2 is valid when $t < 1/\Delta E\sqrt{\sigma}$.
In deriving Eq.D2,
we employ Eq.C2 and notice that
\begin{equation}
M_{l',l}=\delta_{l',l}+\frac{1}{2}[\delta_{l',l+2}+\delta_{l',l-2}]
\end{equation}
in this case.

\end{multicols}

\end{document}